\documentclass[aps,prd,twocolumn,superscriptaddress,amssymb,eqsecnum,showpacs,showkeyes,secnumarabic,graphics,floatfix,nofootinbib,tightenlines,longbibliography]{revtex4-1}
\usepackage{graphicx}
\usepackage{bm}
\usepackage{cancel}
\usepackage{epsfig}
\usepackage{dcolumn}
\usepackage{amsmath}
\usepackage{hhline}
\usepackage{enumerate}
\usepackage[utf8]{inputenc}
\usepackage[dvipsnames]{xcolor}
\usepackage[breaklinks=true,colorlinks=true,
linkcolor=blue,urlcolor=Blue,citecolor=MidnightBlue,
bookmarks=true,bookmarksopenlevel=2]{hyperref}
\usepackage{bm}
\usepackage{caption}
\usepackage{amsmath}
\usepackage {amssymb}
\usepackage{subcaption}

\def \beq  {\begin{equation}}
\def \eeq  {\end{equation}}
\def \ber  {\begin{eqnarray}}
\def \eer  {\end{eqnarray}}
\def \Geff {G_{\rm eff}}

\def \omm  {\Omega_{0 {\rm m}}}

\def \Geff {G_{\rm eff}}

\def \oma {\Omega_{\rm m}}

\begin{document}
\newcommand{\newc}{\newcommand}

\newc{\be}{\begin{equation}}
\newc{\ee}{\end{equation}}
\newc{\ba}{\begin{eqnarray}}
\newc{\ea}{\end{eqnarray}}
\newc{\bea}{\begin{eqnarray*}}
\newc{\eea}{\end{eqnarray*}}
\newc{\D}{\partial}
\newc{\ie}{{\it i.e.} }
\newc{\eg}{{\it e.g.} }
\newc{\etc}{{\it etc.} }
\newc{\etal}{{\it et al.}}
\newc{\lcdm}{$\Lambda$CDM }
\newc{\lcdmnospace}{$\Lambda$CDM}
\newc{\omom}{$\Omega_{0m}$ }
\newc{\plcdm}{Planck15/$\Lambda$CDM }
\newc{\plcdmnospace}{Planck15/$\Lambda$CDM}
\newc{\wlcdm}{WMAP7/$\Lambda$CDM }
\newc{\wlcdmnospace}{WMAP7/$\Lambda$CDM}
\newcommand{\fs}{{\rm{\it f\sigma}}_8}

\newcommand{\nn}{\nonumber}
\newc{\ra}{\Rightarrow}
\title{Evolution of the $f\sigma_8$ tension with the \plcdm determination and implications for modified gravity theories}

\author{Lavrentios Kazantzidis}\email{lkazantzi@cc.uoi.gr}
\affiliation{Department of Physics, University of Ioannina, GR-45110, Ioannina, Greece}
\author{Leandros Perivolaropoulos}\email{leandros@uoi.gr}
\affiliation{Department of Physics, University of Ioannina, GR-45110, Ioannina, Greece}

\date{\today}

\begin{abstract}
We construct an updated extended compilation of  distinct (but possibly correlated) $f\sigma_8(z)$ redshift space distortion (RSD) data published between 2006 and 2018. It consists of 63 data points and is significantly larger than previously used similar data sets. After fiducial model correction we obtain the best fit $\Omega_{0m}-\sigma_8$  \lcdm  parameters and show that they are at a $5\sigma$ tension with the corresponding \plcdm values. Introducing a nontrivial covariance matrix correlating randomly $20\%$ of the RSD data points has no significant effect on the above tension level. We
show that the tension disappears  (becomes less than $1\sigma$) when a subsample of the 20 most recently published data is used.
A partial cause for this reduced tension is the fact that more recent data tend to probe higher redshifts (with higher errorbars) where there is degeneracy among different models due to matter domination.
Allowing for a nontrivial evolution of the effective Newton's constant as $G_{\textrm{eff}}(z)/G_{\textrm{N}} = 1 + g_a \left(\frac{z}{1+z}\right)^2 - g_a \left(\frac{z}{1+z}\right)^4$  ($g_a$ is a parameter) and fixing a \plcdm background we find $g_a=-0.91\pm 0.17$ from the full $f\sigma_8$ data set while the 20 earliest and 20 latest data points imply $g_a=-1.28^{+0.28}_{-0.26}$ and $g_a=-0.43^{+0.46}_{-0.41}$ respectively. Thus, the more recent $f\sigma_8$ data appear to favor GR in contrast to earlier data. Finally, we show that the parametrization $f\sigma_8(z)=\lambda \sigma_8 \Omega(z)^\gamma /(1+z)^\beta$  provides an excellent fit to the solution of the growth equation for both GR ($g_a=0$) and modified gravity ($g_a\neq 0$). 
\end{abstract}
\maketitle

\section{Introduction}
\label{sec:Introduction}

A wide range of different cosmological observations \cite{Percival:2007yw,Alam:2016hwk,Kowalski:2008ez,Riess:2006fw,Jaffe:2000tx,Hicken:2009dk,Sievers:2002tq,Izzo:2015vya,Betoule:2014frx,Efstathiou:2017rgv} are converging to the fact that the expansion rate of the Universe is approximated to high accuracy by the \lcdm model \cite{Carroll:2000fy} as
\be
H(z)^2=H_0^2 \left[\Omega_{0m}(1+z)^3 +(1-\Omega_{0m})\right]\label{eq:hzlcdm}
\ee
where $H_0$ is the Hubble parameter, $z$ is the redshift and $\Omega_{0m}$ is the present matter density parameter. The best fit parameter values for \lcdm  as obtained by Planck \cite{Ade:2015xua} are shown in Table \ref{tab:planck1} and in the context of general relativity (GR) they describe the current concordance model \plcdm. Despite of the consistency of the model with cosmological observations measuring the background expansion rate(eg. Type Ia Supernovae SnIa \cite{Perlmutter:1998np,Riess:1998cb,Suzuki:2011hu,Betoule:2014frx} and baryon acoustic oscillations \cite{Alam:2016hwk,Percival:2007yw}), measurements of the growth rate of cosmological density perturbations have been shown to favor parameter values that are in some tension \cite{Bull:2015stt,Sola:2016zeg,Lin:2017bhs,Lin:2017ikq,Sahni:2014ooa} with the \plcdm parameter values of Table \ref{tab:planck1}. Such probes include weak lensing \cite{Hildebrandt:2016iqg,Joudaki:2017zdt,Jee:2015jta,Abbott:2015swa,Troxel:2017xyo,Kohlinger:2017sxk} and redshift space distortion observations \cite{Nesseris:2017vor,Macaulay:2013swa,Bull:2015stt,Tsujikawa:2015mga,Johnson:2015aaa,Basilakos:2017rgc}. A simple way to account for this tension is to allow \cite{Nesseris:2017vor} for the possibility of extensions of GR in the form of modified theories of gravity \citep{Capozziello:2011et,Cai:2018rzd,Perez-Romero:2017njc,Hu:2007nk,Boubekeur:2014uaa,Gannouji:2008wt,Tsujikawa:2007gd,Masaeli:2016wss,DeFelice:2011hq,Sahni:2002dx,Alam:2016wpf,Novikov:2016hrc,Novikov:2016fzd}.

\begin{table}[b!]
\caption{ \plcdm  parameters from Ref. \cite{Ade:2015xua}. The corresponding \wlcdm from Ref. \cite{Larson:2010gs} are also shown for comparison.}
\label{tab:planck1}
\begin{centering}
\begin{tabular}{ccc}
 \hhline{===}
   \rule{0pt}{3ex} 
  Parameter & \plcdm \cite{Ade:2015xua} &  \wlcdm \cite{Larson:2010gs}\\
    \hline
    \rule{0pt}{3ex}  
$\Omega_b h^2$ & $0.02225\pm0.00016$ & $0.02258  \pm 0.00057$\\
$\Omega_c h^2$ & $0.1198\pm0.0015$  & $0.1109 \pm  0.0056$\\
$n_s$ & $0.9645\pm0.0049$ & $0.963 \pm 0.014$\\
$H_0$ & $67.27\pm0.66$ & $ 71.0 \pm 2.5$\\
$\Omega_{0m}$ & $0.3156\pm0.0091$ &  $0.266 \pm 0.025$\\
$w$ & $-1$  & $-1$\\
$\sigma_8$ & $0.831\pm0.013$  & $0.801 \pm 0.030$\\
\hhline{===}
\end{tabular}
\end{centering}
\end{table}

RSD measurements in galaxy redshift surveys \cite{Blake:2012pj,Jones:2004zy,Alam:2015mbd,Wang:2017wia,Guzzo:2013spa} measure the peculiar velocities of matter and thus infer \cite{Kaiser:1987qv}  the growth rate of cosmological perturbations on a range of redshifts and scales.

Since about 2006 most growth rate measurements are reported as the combination $f(a)\sigma_8(a)$ where $a$ is the scale factor $a=\frac{1}{1+z}$,  $f(a)\equiv d\ln \delta(a)/d\ln a$ is the growth rate of cosmological perturbations, $\delta(a)\equiv \delta \rho/\rho$ is the linear matter overdensity growth factor and $\sigma_8$ is the matter power spectrum normalisation on scales of $8 h^{-1} Mpc$. 

RSDs lead to anisotropies of the power spectrum of perturbations which may lead to the values of $f\sigma_8$ by expanding to Legendre polynomials up to order four and assuming that the true underlying matter power spectrum is isotropic while the anisotropy is due only to the peculiar velocities of galaxies that distort the galaxy distribution in redshift space. 

In practice however the anisotropy of the power spectrum on large scales is not only due to the peculiar galactic velocities but also due to the use of an incorrect fiducial cosmology $H(z)$ assumed in converting the measured angles and redshifts into comoving coordinates in order to construct the correlation function and the corresponding power spectrum \cite{Li:2016wbl,Padmanabhan:2008ag,Macaulay:2013swa}. 
In particular, the comoving distance between a pair of galaxies separated by an angle $d\theta$ is obtained from the Friedmann Robertson Walker (FRW) metric as \cite{Saito16,Wilson:2016ggz,Alam:2015rsa}
\be
d \ell_{\perp} = (1 + z) D_A(z) \, d\theta
\ee
where $D_A(z)$ is the angular diameter distance at the redshift of the pair. Also the corresponding separation along the line of sight is 
\be
d \ell_{\parallel} = \frac{c \ dz}{H(z)}
\ee
where $H(z)$ is the true Hubble expansion rate of the  true underlying cosmology. If a different (fiducial) cosmology $H'(z)$ is assumed instead, the corresponding separations become
\begin{gather}
d \ell_{\perp}' = (1 + z) D_A' d\theta = \left ( \frac{D_A'}{D_A} \right ) \ d \ell_{\perp} =  \frac{d \ell_{\perp}}{f_{\perp}}, \\
d \ell_{\parallel}' = \frac{c \ dz}{H'} = \left ( \frac{H}{H'} \right ) \ d \ell_{\parallel} = \frac{d \ell_{\parallel}}{f_{\parallel}}
\end{gather}
where $F\equiv f_{\parallel}/f_{\perp}$ is the induced anisotropy due to the use of incorrect fiducial cosmology and has magnitude \cite{Li:2016wbl}
\be
F  = \frac{f_{\parallel}}{f_{\perp}} = \left( \frac{H'}{H} \right ) \left( \frac{D_A'}{D_A} \right)
\ee
This induced anisotropy due to the use of incorrect fiducial cosmology is the Alcock-Paczynski (AP) effect \cite{Alcock:1979mp} and is degenerate with the RSD anisotropy induced by the galactic peculiar velocities due to the growth of structures \cite{Alam:2015rsa}. Thus if an  $\fs'$ measurement has been obtained assuming a fiducial \lcdm cosmology $H'(z)$, the corresponding $\fs$ obtained with the true cosmology $H(z)$ is approximated as \cite{,Macaulay:2013swa} 
\be 
f\sigma_8(z)\simeq\frac{H(z) D_A(z)}{H'(z)D_A'(z)} f\sigma_8'(z)\equiv q(z,\Omega_{0m},\Omega_{0m}')\; f\sigma_8'(z)
\label{eq:fs8corr}
\ee
This equation should be taken as a rough order of magnitude estimate of the AP effect as it appears in somewhat different forms in the literature \cite{Saito16,Wilson:2016ggz,Alam:2015rsa}.  In Appendix \ref{sec:Appendix_A} we discuss alternative forms of the correction factor \cite{Wilson:2016ggz,Alam:2015rsa}.

As discussed in Sec. \ref{sec:Section 3}, this correction is small (at most it can be about $2-3\%$ at redshifts $z\simeq 1$ for reasonable values of $\Omega_{0m}$). However we include it in the present analysis and we estimate its effect on the best fit cosmological parameter values.

A compilation of 63 $\fs$ measurements published by various surveys from 2006 to present is shown in Table \ref{tab:all-data} along with the corresponding fiducial cosmology assumed in each case. Despite of the existence of such a large sample of published $\fs$ data, most previous analyses \cite{Nesseris:2017vor,Gomez-Valent:2017idt,Gomez-Valent:2018nib,Basilakos:2017rgc,Basilakos:2016nyg,Nesseris:2014mfa,Mehrabi:2015kta,Baker:2014zva,Nesseris:2007pa,Basilakos:2014yda,Pouri:2014nta,Paul:2013sha,LHuillier:2017ani,Perez-Romero:2017njc,Alam:2015rsa}  that  use growth data to constrain cosmological models use less than 20 data points which are usually selected from the larger data set of Table \ref{tab:all-data} on the basis of subjective criteria that favor more recent data as well as a qualitative minimization of correlations among the selected data points. Indeed, since many of these data points are correlated due to overlap in the galaxy samples used for their derivation, a large covariance matrix should be used for their combined analysis. However no full covariance matrix is available in the literature for the data set of Table \ref{tab:all-data} and for almost all of its subsets. In addition the use of different fiducial models by different surveys at different times is also a source of uncertainty when using large $\fs$ samples. 

Despite these problems, the use of ad hoc subsamples of the full $\fs$ data set of Table \ref{tab:all-data} may lead to a waste of useful information. Therefore, it would be interesting to perform a more detailed analysis of the full $\fs$ sample to identify possible trends of best fit parameters in the context of different subsamples, as well as to study the effects of fiducial cosmology or correlation among data points.

In particular the following open questions are of interest:
\begin{enumerate}[{(1)}]
\item
What is a complete data set of the published $\fs$ data?
\item
What is the tension level of the best fit \lcdm $\Omega_{0m}-\sigma_8$ obtained from the full growth data set with \plcdmnospace?
\item
What is the effect of a typical covariance matrix on the level of the above tension?
\item
Is the tension level the same for early and more recently published RSD $\fs$ data? Is the consistency with GR improving with time of publication of data points?
\item
How is the tension level affected by the $\fs$ correction imposed for the different fiducial cosmologies used in each survey?
\item
Is the spread of the $\fs$ data consistent with the published error bars?
\end{enumerate}

A large part of the present analysis is devoted to the study of these questions. In addition we search for a proper parametrization of $f\sigma_8(z)$ that can represent the predictions of a wide range of cosmological models including models of modified gravity. 

It is well known \cite{Wang:1998gt,Linder:2005in,Polarski:2007rr,Gannouji:2008jr,Polarski:2016ieb,Nesseris:2015fqa} that the growth rate $f(z)$ of cosmological perturbations in the context of general relativity is well approximated by a parametrization of the following form
\ba
&f(a)&=\oma(a)^{\gamma(a)} \label{eq:fg}\\
&\oma(a)& \equiv \frac{\Omega_{0m} ~a^{-3}}{H(a)^2/H_0^2} \\
&\gamma(a)&=\frac{\ln f(a)}{\ln \oma(a)}\simeq 0.55 \label{eq:wst1}
\ea
where we have assumed \lcdm background cosmology. 
\begin{table*}[h!]
\caption{A compilation of RSD data that we found published from 2006 since 2018
\label{tab:all-data}}
\begin{centering}
\begin{tabular}{ccccccc}
\hhline{=======}
  \rule{0pt}{3ex}  
Index & Dataset & $z$ & $f\sigma_8(z)$ & Refs. & Year & Fiducial Cosmology \\
\hline
1 & SDSS-LRG & $0.35$ & $0.440\pm 0.050$ & \cite{Song:2008qt} &  30 October 2006 &$(\Omega_{0m},\Omega_K,\sigma_8$)$=(0.25,0,0.756)$\cite{Tegmark:2006az} \\

2 & VVDS & $0.77$ & $0.490\pm 0.18$ & \cite{Song:2008qt}  & 6 October 2009 & $(\Omega_{0m},\Omega_K,\sigma_8)=(0.25,0,0.78)$ \\

3 & 2dFGRS & $0.17$ & $0.510\pm 0.060$ & \cite{Song:2008qt}  &  6 October 2009 & $(\Omega_{0m},\Omega_K)=(0.3,0,0.9)$ \\

4 & 2MRS &0.02& $0.314 \pm 0.048$ &  \cite{Davis:2010sw}, \cite{Hudson:2012gt}& 13 Novemver 2010 & $(\Omega_{0m},\Omega_K,\sigma_8)=(0.266,0,0.65)$ \\

5 & SnIa+IRAS &0.02& $0.398 \pm 0.065$ & \cite{Turnbull:2011ty}, \cite{Hudson:2012gt}& 20 October 2011 & $(\Omega_{0m},\Omega_K,\sigma_8)=(0.3,0,0.814)$\\

6 & SDSS-LRG-200 & $0.25$ & $0.3512\pm 0.0583$ & \cite{Samushia:2011cs} & 9 December 2011 & $(\Omega_{0m},\Omega_K,\sigma_8)=(0.276,0,0.8)$  \\

7 & SDSS-LRG-200 & $0.37$ & $0.4602\pm 0.0378$ & \cite{Samushia:2011cs} & 9 December 2011 & \\

8 & SDSS-LRG-60 & $0.25$ & $0.3665\pm0.0601$ & \cite{Samushia:2011cs} & 9 December 2011 & $(\Omega_{0m},\Omega_K,\sigma_8)=(0.276,0,0.8)$ \\

9 & SDSS-LRG-60 & $0.37$ & $0.4031\pm0.0586$ & \cite{Samushia:2011cs} & 9 December 2011 &\\

10 & WiggleZ & $0.44$ & $0.413\pm 0.080$ & \cite{Blake:2012pj} & 12 June 2012  & $(\Omega_{0m},h,\sigma_8)=(0.27,0.71,0.8)$ \\

11 & WiggleZ & $0.60$ & $0.390\pm 0.063$ & \cite{Blake:2012pj} & 12 June 2012 &  $C_{ij}=Eq.\eqref{eq:wigglez}$\\

12 & WiggleZ & $0.73$ & $0.437\pm 0.072$ & \cite{Blake:2012pj} & 12 June 2012 &\\

13 & 6dFGS& $0.067$ & $0.423\pm 0.055$ & \cite{Beutler:2012px} & 4 July 2012 & $(\Omega_{0m},\Omega_K,\sigma_8)=(0.27,0,0.76)$ \\

14 & SDSS-BOSS& $0.30$ & $0.407\pm 0.055$ & \cite{Tojeiro:2012rp} & 11 August 2012 & $(\Omega_{0m},\Omega_K,\sigma_8)=(0.25,0,0.804)$ \\

15 & SDSS-BOSS& $0.40$ & $0.419\pm 0.041$ & \cite{Tojeiro:2012rp} & 11 August 2012 & \\

16 & SDSS-BOSS& $0.50$ & $0.427\pm 0.043$ & \cite{Tojeiro:2012rp} & 11 August 2012 & \\

17 & SDSS-BOSS& $0.60$ & $0.433\pm 0.067$ & \cite{Tojeiro:2012rp} & 11 August 2012 & \\

18 & Vipers& $0.80$ & $0.470\pm 0.080$ & \cite{delaTorre:2013rpa} & 9 July 2013 & $(\Omega_{0m},\Omega_K,\sigma_8)=(0.25,0,0.82)$  \\

19 & SDSS-DR7-LRG & $0.35$ & $0.429\pm 0.089$ & \cite{Chuang:2012qt}  & 8 August 2013 & $(\Omega_{0m},\Omega_K,\sigma_8$)$=(0.25,0,0.809)$\cite{Komatsu:2010fb}\\

20 & GAMA & $0.18$ & $0.360\pm 0.090$ & \cite{Blake:2013nif}  & 22 September 2013 & $(\Omega_{0m},\Omega_K,\sigma_8)=(0.27,0,0.8)$ \\

21& GAMA & $0.38$ & $0.440\pm 0.060$ & \cite{Blake:2013nif}  & 22 September 2013 & \\

22 & BOSS-LOWZ& $0.32$ & $0.384\pm 0.095$ & \cite{Sanchez:2013tga}  & 17 December 2013  & $(\Omega_{0m},\Omega_K,\sigma_8)=(0.274,0,0.8)$ \\

23 & SDSS DR10 and DR11 & $0.32$ & $0.48 \pm 0.10$ & \cite{Sanchez:2013tga} &   17 December 2013 & $(\Omega_{0m},\Omega_K,\sigma_8$)$=(0.274,0,0.8)$\cite{Anderson:2013zyy}\\

24 & SDSS DR10 and DR11 & $0.57$ & $0.417 \pm 0.045$ & \cite{Sanchez:2013tga} &  17 December 2013 &  \\

25 & SDSS-MGS & $0.15$ & $0.490\pm0.145$ & \cite{Howlett:2014opa} & 30 January 2015 & $(\Omega_{0m},h,\sigma_8)=(0.31,0.67,0.83)$ \\

26 & SDSS-veloc & $0.10$ & $0.370\pm 0.130$ & \cite{Feix:2015dla}  & 16 June 2015 & $(\Omega_{0m},\Omega_K,\sigma_8$)$=(0.3,0,0.89)$\cite{Tegmark:2003uf} \\

27 & FastSound& $1.40$ & $0.482\pm 0.116$ & \cite{Okumura:2015lvp}  & 25 November 2015 & $(\Omega_{0m},\Omega_K,\sigma_8$)$=(0.27,0,0.82)$\cite{Hinshaw:2012aka} \\

28 & SDSS-CMASS & $0.59$ & $0.488\pm 0.060$ & \cite{Chuang:2013wga} & 8 July 2016 & $\ \ (\Omega_{0m},h,\sigma_8)=(0.307115,0.6777,0.8288)$ \\

29 & BOSS DR12 & $0.38$ & $0.497\pm 0.045$ & \cite{Alam:2016hwk} & 11 July 2016 & $(\Omega_{0m},\Omega_K,\sigma_8)=(0.31,0,0.8)$ \\

30 & BOSS DR12 & $0.51$ & $0.458\pm 0.038$ & \cite{Alam:2016hwk} & 11 July 2016 & \\

31 & BOSS DR12 & $0.61$ & $0.436\pm 0.034$ & \cite{Alam:2016hwk} & 11 July 2016 & \\

32 & BOSS DR12 & $0.38$ & $0.477 \pm 0.051$ & \cite{Beutler:2016arn} & 11 July 2016 & $(\Omega_{0m},h,\sigma_8)=(0.31,0.676,0.8)$ \\

33 & BOSS DR12 & $0.51$ & $0.453 \pm 0.050$ & \cite{Beutler:2016arn} & 11 July 2016 & \\

34 & BOSS DR12 & $0.61$ & $0.410 \pm 0.044$ & \cite{Beutler:2016arn} & 11 July 2016 &  \\

35 &Vipers v7& $0.76$ & $0.440\pm 0.040$ & \cite{Wilson:2016ggz} & 26 October 2016  & $(\Omega_{0m},\sigma_8)=(0.308,0.8149)$ \\

36 &Vipers v7 & $1.05$ & $0.280\pm 0.080$ & \cite{Wilson:2016ggz} & 26 October 2016 &\\

37 &  BOSS LOWZ & $0.32$ & $0.427\pm 0.056$ & \cite{Gil-Marin:2016wya} & 26 October 2016 & $(\Omega_{0m},\Omega_K,\sigma_8)=(0.31,0,0.8475)$\\

38 & BOSS CMASS & $0.57$ & $0.426\pm 0.029$ & \cite{Gil-Marin:2016wya} & 26 October 2016 & \\

39 & Vipers  & $0.727$ & $0.296 \pm 0.0765$ & \cite{Hawken:2016qcy} &  21 November 2016 & $(\Omega_{0m},\Omega_K,\sigma_8)=(0.31,0,0.7)$\\

40 & 6dFGS+SnIa & $0.02$ & $0.428\pm 0.0465$ & \cite{Huterer:2016uyq} & 29 November 2016 & $(\Omega_{0m},h,\sigma_8)=(0.3,0.683,0.8)$ \\

41 & Vipers  & $0.6$ & $0.48 \pm 0.12$ & \cite{delaTorre:2016rxm} & 16 December 2016 & $(\Omega_{0m},\Omega_b,n_s,\sigma_8$)= $(0.3, 0.045, 0.96,0.831)$\cite{Ade:2015xua} \\

42 & Vipers  & $0.86$ & $0.48 \pm 0.10$ & \cite{delaTorre:2016rxm} & 16 December 2016  & \\

43 &Vipers PDR-2& $0.60$ & $0.550\pm 0.120$ & \cite{Pezzotta:2016gbo} & 16 December 2016 & $(\Omega_{0m},\Omega_b,\sigma_8)=(0.3,0.045,0.823)$ \\

44 & Vipers PDR-2& $0.86$ & $0.400\pm 0.110$ & \cite{Pezzotta:2016gbo} & 16 December 2016 &\\

45 & SDSS DR13  & $0.1$ & $0.48 \pm 0.16$ & \cite{Feix:2016qhh} & 22 December 2016 & $(\Omega_{0m},\sigma_8$)$=(0.25,0.89)$\cite{Tegmark:2003uf} \\

46 & 2MTF & 0.001 & $0.505 \pm 0.085$ &  \cite{Howlett:2017asq} & 16 June 2017 & $(\Omega_{0m},\sigma_8)=(0.3121,0.815)$\\

47 & Vipers PDR-2 & $0.85$ & $0.45 \pm 0.11$ & \cite{Mohammad:2017lzz} & 31 July 2017  &  $(\Omega_b,\Omega_{0m},h)=(0.045,0.30,0.8)$ \\

48 & BOSS DR12 & $0.31$ & $0.469 \pm 0.098$ &  \cite{Wang:2017wia} & 15 September 2017 & $(\Omega_{0m},h,\sigma_8)=(0.307,0.6777,0.8288)$\\

49 & BOSS DR12 & $0.36$ & $0.474 \pm 0.097$ &  \cite{Wang:2017wia} & 15 September 2017 & \\

50 & BOSS DR12 & $0.40$ & $0.473 \pm 0.086$ &  \cite{Wang:2017wia} & 15 September 2017 & \\

51 & BOSS DR12 & $0.44$ & $0.481 \pm 0.076$ &  \cite{Wang:2017wia} & 15 September 2017 & \\

52 & BOSS DR12 & $0.48$ & $0.482 \pm 0.067$ &  \cite{Wang:2017wia} & 15 September 2017 & \\

53 & BOSS DR12 & $0.52$ & $0.488 \pm 0.065$ &  \cite{Wang:2017wia} & 15 September 2017 & \\

54 & BOSS DR12 & $0.56$ & $0.482 \pm 0.067$ &  \cite{Wang:2017wia} & 15 September 2017 & \\

55 & BOSS DR12 & $0.59$ & $0.481 \pm 0.066$ &  \cite{Wang:2017wia} & 15 September 2017 & \\

56 & BOSS DR12 & $0.64$ & $0.486 \pm 0.070$ &  \cite{Wang:2017wia} & 15 September 2017 & \\

57 & SDSS DR7 & $0.1$ & $0.376\pm 0.038$ & \cite{Shi:2017qpr} & 12 December 2017 & $(\Omega_{0m},\Omega_b,\sigma_8)=(0.282,0.046,0.817)$ \\

58 & SDSS-IV & $1.52$ & $0.420 \pm 0.076$ &  \cite{Gil-Marin:2018cgo} & 8 January 2018  & $(\Omega_{0m},\Omega_b h^2,\sigma_8)=(0.26479, 0.02258,0.8)$ \\ 

59 & SDSS-IV & $1.52$ & $0.396 \pm 0.079$ & \cite{Hou:2018yny} & 8 January 2018 & $(\Omega_{0m},\Omega_b h^2,\sigma_8)=(0.31,0.022,0.8225)$ \\ 

60 & SDSS-IV & $0.978$ & $0.379 \pm 0.176$ &  \cite{Zhao:2018jxv} & 9 January 2018 &$(\Omega_{0m},\sigma_8)=(0.31,0.8)$\\

61 & SDSS-IV & $1.23$ & $0.385 \pm 0.099$ &  \cite{Zhao:2018jxv} & 9 January 2018 & \\

62 & SDSS-IV & $1.526$ & $0.342 \pm 0.070$ &  \cite{Zhao:2018jxv} & 9 January 2018 & \\

63 & SDSS-IV & $1.944$ & $0.364 \pm 0.106$ &  \cite{Zhao:2018jxv} & 9 January 2018 & \\
\hhline{=======}
\end{tabular}\par\end{centering}
\end{table*}

\newpage

The construction of a  corresponding parametrization that approximates well the theoretically predicted form of $\fs(z)$ for a wide range of theoretical models is an interesting open question that is addressed in the present analysis.

The structure of this paper is the following: In the next section we review the equations that determine the growth of matter perturbations in GR and in modified gravity theories as parametrized by the effective Newton's constant $G_{eff}$. We compare the numerical solution for $\fs(z)$ in the context of different cosmological models and present a new parametrization for $\fs(z)$ which provides an excellent fit to the numerical solution of $\fs$ for both \lcdm and modified gravity models. This parametrization may be viewed as an extension for the corresponding parametrization Eq. \eqref{eq:fg} for the growth rate $f(a)$. In Sec. \ref{sec:Section 3} we present a detailed analysis of the data set of Table \ref{tab:all-data} addressing the questions stated above using appropriate statistics. Finally in Sec. \ref{sec:Section 4} we summarise and discuss implications and future prospects of our results.

\section{Theoretical Predictions of $f\sigma_8(z)$ }
The \plcdm concordance background model described by Eq. \eqref{eq:hzlcdm} with parameters from Table \ref{tab:planck1} can be reproduced by a wide range of theoretical models including models with dynamical and/or clustering dark energy and modified gravity models. In order to efficiently discriminate among these classes of models, the evolution of matter density perturbations must be considered and its theoretically predicted evolution must be compared with cosmological observations. The equation that describes the evolution of the linear matter  growth factor $\delta\equiv \delta \rho/\rho$ in the context of both GR and most modified gravity theories is of the form
\be
{\ddot \delta} + 2H {\dot \delta} - 4\pi G_{\rm eff}\,
\rho \, \delta \approx 0 \label{eq:odedeltat}
\ee
where $\rho$ is the background matter density and  $G_{\rm eff}$ is the effective Newton's constant which in general may depend on both redshift $z$ and cosmological scale $k$. Eq. \eqref{eq:odedeltat} in terms of the redshift $z$  takes the following form 
\begin{equation}
\delta'' + \left(\frac{(H^2)'}{2~H^2} -
{1\over 1+z}\right)\delta'
\approx {3\over 2} (1+z)\frac{ H_0^2}{H^2} {G_{\rm eff}(z,k)\over
G_{N}}~\Omega_{0m}\delta
\label{eq:odedeltaz}
\end{equation}
The effective Newton's constant  arises from a generalized Poisson equation of the following form
\be 
\nabla^2 \phi \approx 4 \pi G_{\rm eff} \rho \; \delta,
\ee
where $\phi$ is the perturbed metric potential in the Newtonian gauge defined via the perturbed FRW metric
\be
ds^2= -(1 + 2 \phi) dt^2 + a^2 (1 - 2\psi) d{\vec{x}}\,^2
\ee
 In GR we have a constant homogeneous $G_{\rm eff}(z,k)=G_N$ ($G_N$ is Newton's constant as measured by local experiments) while in modified gravity theories $G_{\rm eff}/G_N$ may vary in both cosmological times (redshifts) and scales. In terms of the scale factor instead of redshift, Eq. \eqref{eq:odedeltaz} may be expressed as
\be
\delta''(a)+\left(\frac{3}{a}+\frac{H'(a)}{H(a)}\right)\delta'(a)
-\frac{3}{2}\frac{\Omega_{0m} \Geff(a,k)/G_{\textrm{N}}}{a^5 H(a)^2/H_0^2}~\delta(a)=0 \label{eq:odedeltaa}
\ee

For example in a modified gravity theory with action of the form
\be
S=\int d^4x \sqrt{-g}\left(\frac12f(R,\phi,X)+\mathcal{L}_m\right),\label{eq:mogaction}
\ee
where $X=-g^{\mu\nu}\partial_\mu \phi \partial_\nu \phi$,   $G_{\rm eff}$ is expressed as 
\be
\Geff(a,k)/G_{\textrm{N}}=\frac{1}{F}\frac{f_{,X}+4\left(f_{,X} \frac{k^2}{a^2}\frac{F_{,R}}{F}+\frac{F_{,\phi}^2}{F}\right)}{f_{,X}+3\left(f_{,X} \frac{k^2}{a^2}\frac{F_{,R}}{F}+\frac{F_{,\phi}^2}{F}\right)},\label{eq:geffmog}
\ee
where $F=F(R,\phi,X)=\partial_R f(R,\phi,X)$ and $F_{,\phi}=\partial_\phi F(R,\phi,X)$. For scalar-tensor theories \cite{Lykkas:2015kls,Tsujikawa:2007gd,Boisseau:2000pr,EspositoFarese:2000ij,Perivolaropoulos:2005yv} with Lagrangian density 
\be
\mathcal{L}^{\textrm{ScT}}=\frac{F(\phi)}{2}R+X-U(\phi) \label{eq:SCTEN}
\ee
$G_{\rm eff}/G_N$ takes the form
\be
\Geff(a,k)/G_{\textrm{N}}=\frac{1}{F(\phi)}\frac{F(\phi)+2F_{,\phi}^2}
{F(\phi)+\frac32F_{,\phi}^2}.\label{eq:geffSCT}
\ee
Solar system tests impose the following constraint on $G_{\rm eff}$  \cite{Nesseris:2006hp} 
\be 
\Big\lvert \frac{1}{G_N} \frac{d G_{\rm eff}(z)}{dz} \Big \vert_{z=0} \Big\rvert < 10^{-3} h^{-1}
\label{eq:geffconstr1}
\ee
while the second derivative is effectively unconstrained since \cite{Nesseris:2006hp} 
\be 
\Big\vert \frac{1}{G_N}  \frac{d^2G_{\rm eff}(z)}{dz^2} \Big \vert_{z=0}\Big\vert < 10^{5} h^{-2}
\label{eq:geffconstr2}
\ee
In addition, nucleosynthesis constraints \cite{Copi:2003xd} imply that at $1\sigma$ 
\be
\lvert G_{\rm eff}/G_{N} -1 \rvert \leq 0.2
\label{eq:geffnucconstr}
\ee

These constraints are respected by a parametrization of $G_{\rm eff}(z)$ of the form \cite{Nesseris:2017vor}
\ba
\frac{G_{\textrm{eff}}(a,g_a,n)}{G_{\textrm{N}}} &=& 1+g_a(1-a)^n - g_a(1-a)^{n+m} \nn \\
&=&1+g_a\left(\frac{z}{1+z}\right)^n - g_a\left(\frac{z}{1+z}\right)^{n+m}. \label{eq:geffansatz}
\ea
where $n,m$ are integer parameters with $n\geq 2$ and $m>0$. In what follows we set $n=m=2$. For these parameter values, the parameter $g_a$ is constrained by the integrated
Sachs-Wolfe effect from the CMB power spectrum to be $g_a< 0.5$ \cite{Nesseris:2017vor}.

The observable quantity $f\sigma_8(a)$ can be derived from the solution $\delta(a)$ of Eq. \eqref{eq:odedeltaa} using the definitions $f(a)\equiv d\ln \delta(a)/ d\ln a$ and $\sigma(a)=\sigma_8 \frac{\delta(a)}{\delta{1}}$. Hence \cite{Percival:2005vm}

\ba
\fs(a)&\equiv& f(a)\cdot \sigma(a)=\frac{\sigma_8}{\delta(1)}~a~\delta'(a) ,\label{eq:fs8}
\ea

\begin{figure}[!h]
\centering
\includegraphics[width = 0.48\textwidth]{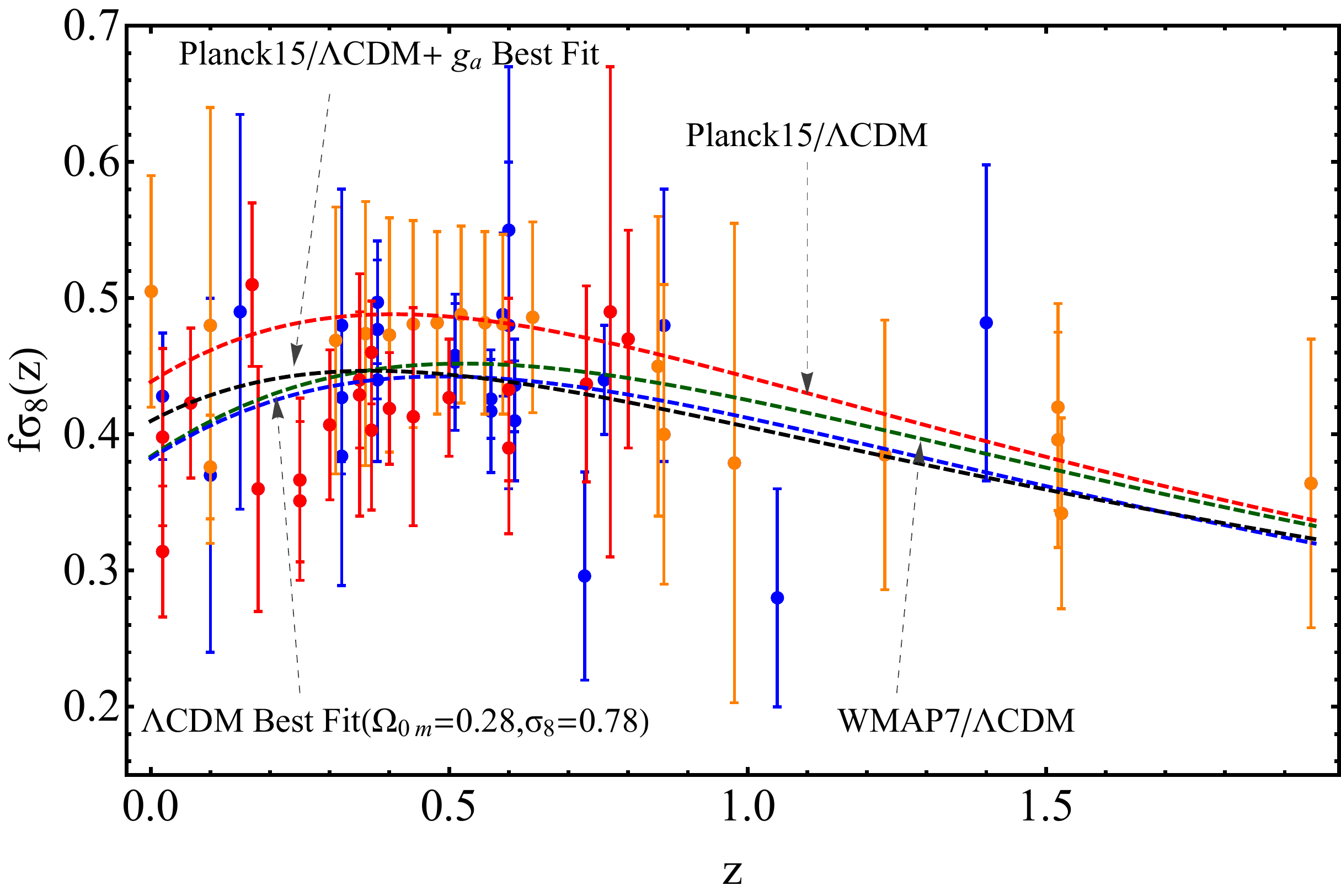}
\caption{Plot of $f\sigma_8(z)$ for the full growth rate data set. The green dashed line and the red dashed one correspond to the best fits of \wlcdm and \plcdm models respectively whereas the blue one describes the best fit \lcdm ($\Omega_{0m}=0.28\pm 0.02$) to the full growth data set and the black one to the \plcdm with $g_a$ best fit. The red points correspond to the 20 earliest published points whereas the orange ones to the 20 latest published points taking into account Table \ref{tab:all-data}.}
\label{fig:fs8z}
\end{figure}

Therefore the prediction of $f\sigma_8(a)$ [or equivalently $f\sigma_8(z)$] is obtained by solving numerically Eq. \eqref{eq:fs8}\footnote{There are analytic solutions of Eq. \eqref{eq:fs8} expressed in terms of hyperheometric functions for specific cosmological models including \lcdm\cite{Nesseris:2015fqa,Belloso:2011ms,Silveira:1994yq,Percival:2005vm}.} in the range $a\in [0,1]$ with initial conditions assuming GR and matter domination  (we set initially $\delta(a)\simeq a$) and using Eq.\eqref{eq:fs8}. The $f\sigma_8(z)$ solutions for a \plcdm and for \wlcdm background cosmology $H(z)$ are shown in Fig.  \ref{fig:fs8z} along with the data of Table \ref{tab:all-data}. 

Notice that \wlcdm appears to be more consistent with the full $\fs$ data set than \plcdm which appears to predict a larger $\fs$ than favored by the data. This well known tension will be analysed in detail in the next section.

\begin{figure}[!h]
\centering
\includegraphics[width = 0.48\textwidth]{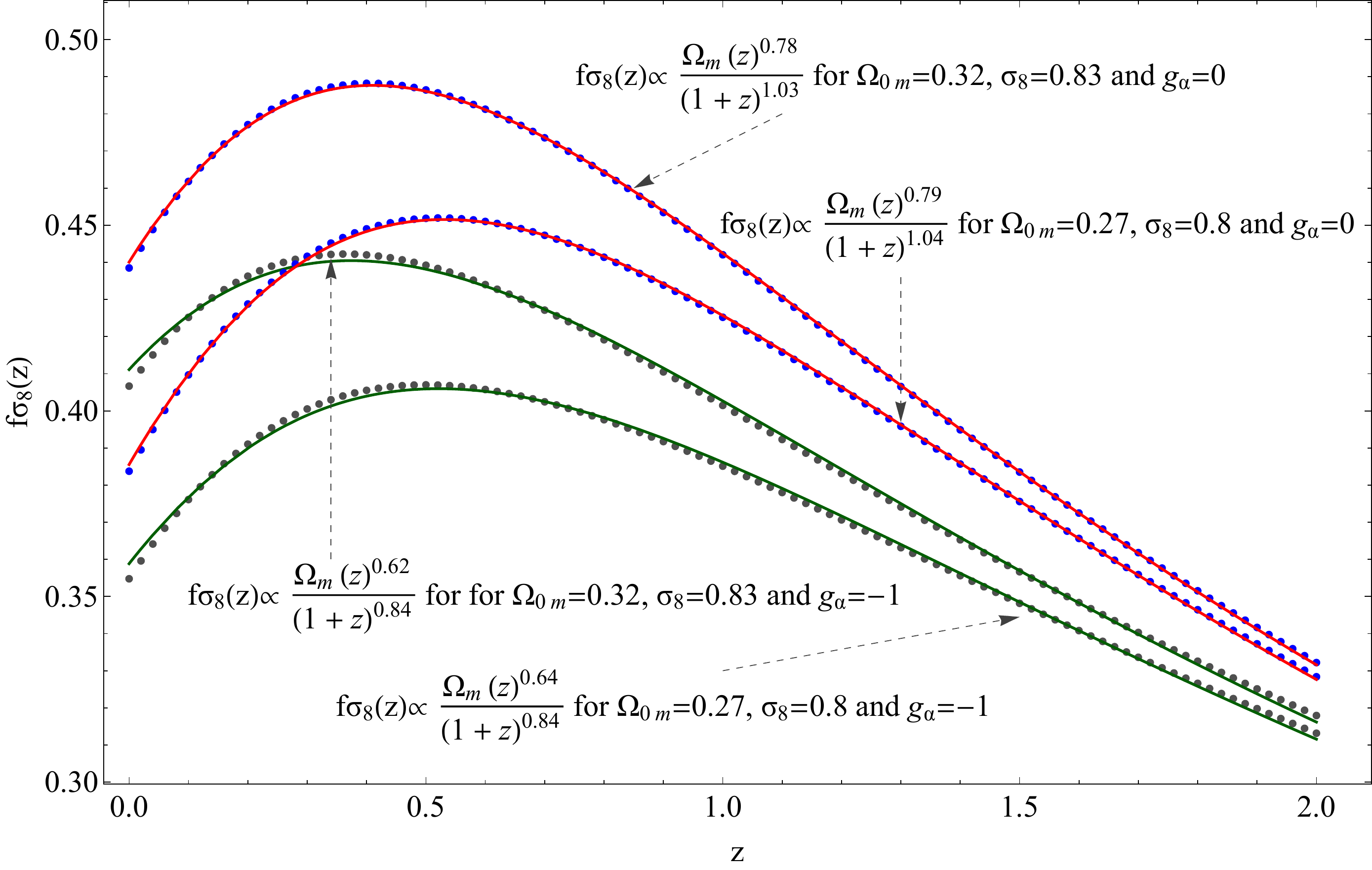}
\caption{Parametrization \eqref{eq:fs8param} for \plcdm and \wlcdm. The thick blue dots  of the upper (lower) curve correspond to the  numerical solution of $\fs(z)$ for \plcdm (\wlcdmnospace) superposed with the analytic form \eqref{eq:fs8param} (red lines) assuming GR, whereas the gray ones of the upper (lower) curve represent the  numerical solution of $\fs(z)$ for \plcdm (\wlcdmnospace) superposed with the analytic form \eqref{eq:fs8param} (green lines) for modified gravity, \ie $g_a=-1$.}
\label{fig:numericsol}
\end{figure}

Even though there are analytic solutions to Eq. \eqref{eq:odedeltaa} expressed in terms of hypergeometric functions \cite{Silveira:1994yq,Percival:2005vm,Belloso:2011ms,Nesseris:2015fqa} it would be useful to provide a parametrization for $f\sigma_8 (z)$ in analogy with the $f(z)$ parametrization of Eq. \eqref{eq:fg}. In view of the fact that $\sigma_8(a)\sim \delta(a)$ while $\delta(a)\sim a =\frac{1}{1+z}$ in a flat matter dominated universe, it is natural to anticipate a parametrization of the form
\be
f\sigma_8(z)= \lambda \sigma_8 \frac{\Omega_m(z)^\gamma}{(1+z)^\beta}\label{eq:fs8param}
\ee
where
\be
\Omega_m(z)= \frac{\omm  (z+1)^3}{\omm (z+1)^3+1-\omm}\label{eq:omegazparam}
\ee
and $\lambda$, $\beta$, $\gamma$ are parameters to be determined for given cosmological model. The parametrization \eqref{eq:fs8param} provides an excellent fit to the numerical solution $f\sigma_8(z)$. This is demonstrated in Fig. \ref{fig:numericsol} where we show the numerical solution for $f\sigma_8(z)$ (dotted lines) for \plcdm and \wlcdm (GR is assumed $g_a=0$) superposed with the analytic form \eqref{eq:omegazparam} (continous red lines)  for $\gamma\simeq 0.78$ and $\beta \simeq 1$ (the exact parameter values are shown on the figure caption for each case). 

Similarly, under the assumption of modified gravity ($g_a=-1$),\footnote{This value for $g_a$ is motivated from the analysis of Ref. \cite{Nesseris:2017vor} that indicated that such a value of $g_a$ can reduce the tension between the $\fs$ data and a \plcdm $H(z)$ background.} the numerical solution (dotted lines) is shown in the same figure for the same backgrounds $H(z)$ superposed with the corresponding analytic parametrization (continuous green lines). 

\begin{figure*}[!t]
\centering
\begin{center}
$\begin{array}{@{\hspace{-0.10in}}c@{\hspace{0.0in}}c}
\multicolumn{1}{l}{\mbox{}} &
\multicolumn{1}{l}{\mbox{}} \\ [-0.2in]
\epsfxsize=3.3in
\epsffile{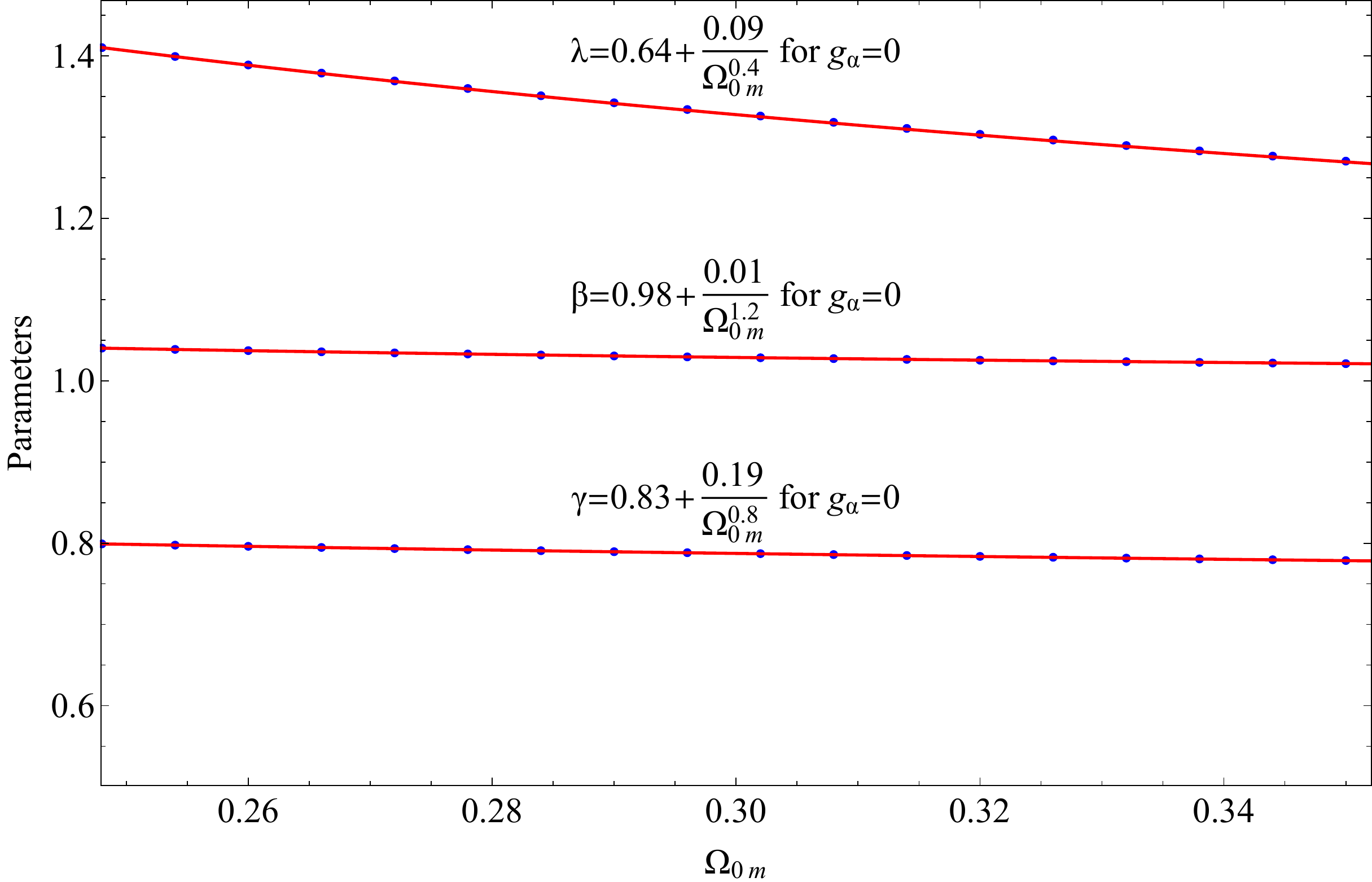} &
\epsfxsize=3.3in
\epsffile{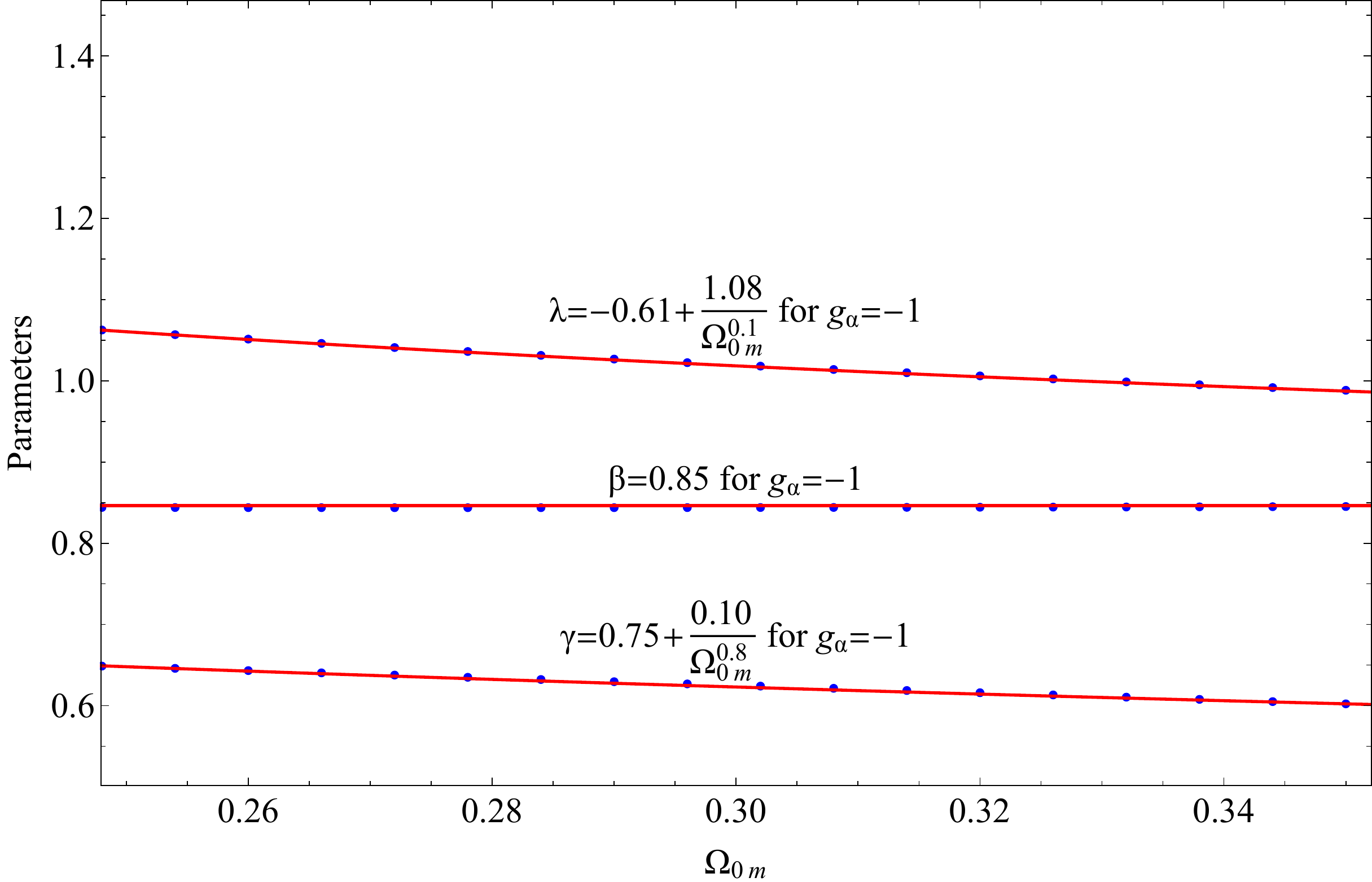} \\
\end{array}$
\end{center}
\vspace{0.0cm}
\caption{The dependence of the parameters $\lambda$ (upper curve), $\beta$ (middle curve) and $\gamma$ (lower curve) on \omom. The  blue dots are the numerically obtained values while the continuous (red) lines correspond to the best fit  power laws for GR (left figure) and modified gravity (right figure), \ie $g_a=-1$.  In the two plots we use the same range for comparison.}
\label{fig:powerlaws}
\end{figure*}

The parametrization \eqref{eq:fs8param} continues to provide still an excellent fit but for somewhat lower values of the parameters ($\beta\simeq 0.84$, $\gamma \simeq 0.63$). Next, in Fig. \ref{fig:powerlaws}, we show the dependence of the parameters $\lambda$, $\beta$ $\gamma$ on $\Omega_{0m}$ for $g_a=0$ and $g_a=-1$.  The dots are numerically obtained values and the continuous lines are power laws that describe the dependence of the parameters on $\Omega_{0m}$. In the range $\Omega_{0m}\in [0.25,0.35]$ and assuming GR ($g_a=0$) we have  $\gamma=0.78 \pm 0.01$, $\lambda=1.3 \pm 0.1$,  $\beta=1.03 \pm 0.01$.  

\section{Consistency of RSD data with \plcdm: Trends and statistics}
\label{sec:Section 3}
\subsection{Trends and Inhomogeneities in the $f\sigma_8$ data}

The full RSD $\fs$ data set of Table \ref{tab:all-data} could be used directly to identify the best fit form of the background cosmology $H(z)$ and/or the best fit form of $G_{\rm eff}(z)$ using the numerical solution of Eq. \eqref{eq:odedeltaa} to construct the predicted $f\sigma_8(z)$ with Eq. \eqref{eq:fs8} and fitting it to the data of Table \ref{tab:all-data}. The results of such a brute force approach should be interpreted with care as they are affected by three factors that may lead to misleading results
\begin{enumerate}[{(1)}]
\item
Correlations Among Data Points: As mentioned in the Introduction, the covariance matrix for the data points of Table \ref{tab:all-data} is not known. This is a source of uncertainty when fitting cosmological models to either the full set of data or to subsets of it.
\item
Fiducial Model Correction: The different fiducial cosmologies for $\fs$ data points shown in Table \ref{tab:all-data} introduce another source of uncertainty that needs to be taken into account when estimating the tension with \plcdmnospace. A proper account of this effect would require a full reconstruction of the correlation function under a \plcdm fiducial cosmology for all data points of Table \ref{tab:all-data}. Alternatively, an approximate correction would be to include an AP correction factor, \ie Eq. \eqref{eq:fs8corr}. 
\item
Survey Systematics: Systematics of surveys that may vary with time of publication and may lead to data inhomogeneities.
\end{enumerate}
In this section we estimate the magnitude of these effects on the tension level of the full $\fs$ data set with \plcdm and on the best fit values of the parameters $\Omega_{0m}-\sigma_8$.

We use the full $\fs$ data set of Table \ref{tab:all-data} to obtain the best fit $\Omega_{0m}-\sigma_8$ parameters in the context of a \lcdm background using the maximum likelihood method. Our method involves the following steps:
\begin{enumerate}[(i)]
\item
Solve Eq. \eqref{eq:odedeltaa} numerically and using Eq. \eqref{eq:fs8} obtain $f\sigma_8(z,\Omega_{0m},\sigma_8,g_a)$ assuming a \lcdm backdound. In this subsection we consider GR and set $g_a=0$ but in the next subsection  we consider also a $G_{\rm eff}$ that is allowed to have a redshift dependence in accordance with the parametrization (\ref{eq:geffansatz}).
\item
Multiply the $\fs$ data of Table \ref{tab:all-data} (and their errorbars) by the fiducial correction factor $q(z,\Omega_{0m},\Omega_{0m}^{fid})=\frac{H(z)D_A(z)}{H^{fid}(z)D_A^{fid}(z)}$ in accordance with Eq. \eqref{eq:fs8corr} where the denominator is obtained from the fiducial \lcdm model of each survey and the numerator involves the $\Omega_{0m}$ parameter to be fit. In practice this factor differs from unity by not more than $2-3\%$ and thus as it will be seen below it does not affect the tension between \plcdm and the growth best fit \lcdm model.
\begin{figure*}[ht!]
\centering
\includegraphics[width = 1.\textwidth]{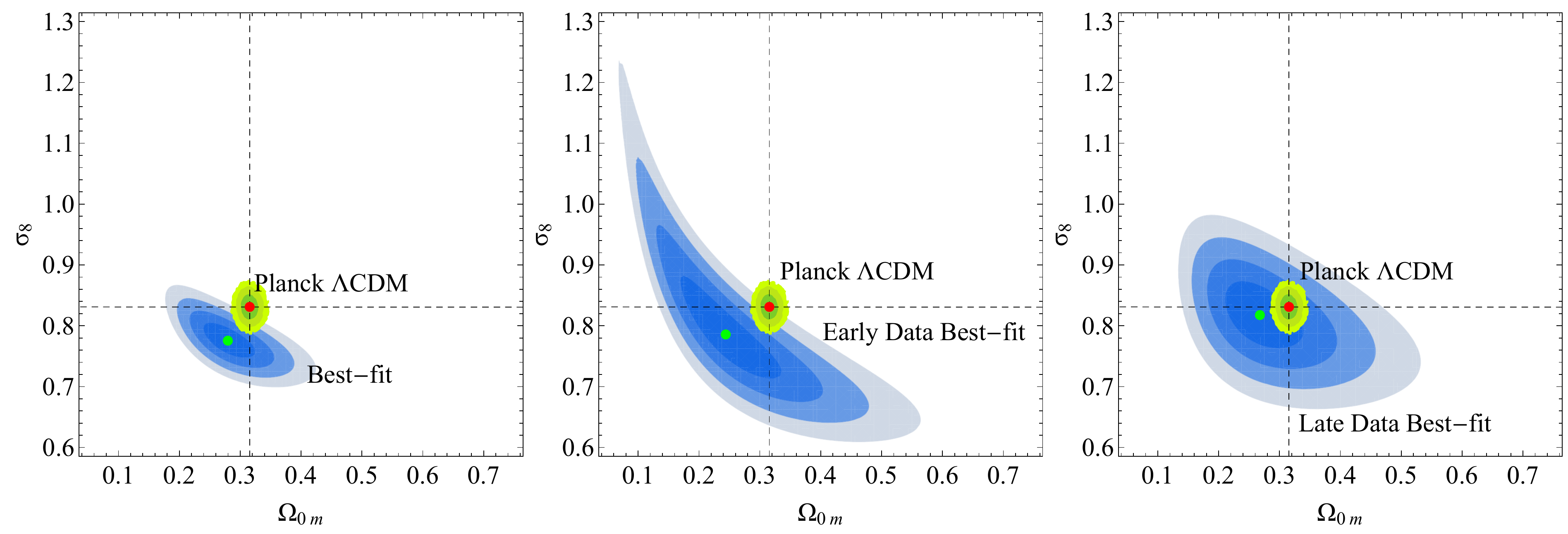}
\caption{The $1\sigma-4\sigma$ confidence contours in the $(\Omega_{0m}-\sigma_8)$ parametric space. The blue contours correspond to the best fit of the 63 compilation data (left panel), the 20 early data (middle panel) and the 20 late data (right panel). The light green contours to the \plcdm, while the red and green dots correspond to the \plcdm best-fit cosmology and the best fit from the growth data respectively.}
\label{fig:fullcontour}
\end{figure*}
\item
As a first step for the construction of $\chi^2$ to be minimized, construct the vector 
\ba
V^i(z_i,\Omega_{0m},\sigma_8,g_a)\equiv f\sigma_{8i} -\frac{ f\sigma_8 (z_i,\Omega_{0m},\sigma_8,g_a)}{q(z,\Omega_{0m},\Omega_{0m}^{fid_i})}\nn\\
\label{eq:vecvidef}
\ea
where we have divided the theoretical prediction  $f\sigma_8 (z_i,\Omega_{0m},\sigma_8,g_a)$ by the correction factor $q$ instead of the equivalent multiplication of the data point $f\sigma_{8i}$ (and its errorbar) by the same correction factor $q$.
\item
Construct the $\chi^2$ to be minimized as
\be
\chi^2 = V^i C_{ij}^{-1}V^j 
\label{chi2}
\ee
where $C_{ij}^{-1}$ is the inverse covariance matrix. We assume that the covariance matrix is diagonal except of the WiggleZ subset of the data (three data points) where the covariance matrix has been published as

\be
C_{ij}^{\text{WiggleZ}}=10^{-3}\left(
         \begin{array}{ccc}
           6.400 & 2.570 & 0.000 \\
           2.570 & 3.969 & 2.540 \\
           0.000 & 2.540 & 5.184 \\
         \end{array}
       \right) \label{eq:wigglez}
\ee
Notice that the $C_{ij}$ nondiagonal element of the WiggleZ covariance matrix is well approximated as $C_{ij}\simeq 0.5 \sqrt{C_{ii}C_{jj}}$. We use this approximation in what follows for the construction of Monte Carlo correlations among the $f\sigma_8$ data points in order to estimate the effects of the ignored correlations among the other data points.
Thus the total covariance matrix takes the form
\be
C_{ij}^{\textrm{growth,total}}=\left(
         \begin{array}{cccc}
           \sigma_1^2 & 0 & 0 & \cdots \\
           0 & C_{ij}^{WiggleZ} & 0& \cdots \\
           0 & 0 & \cdots &   \sigma_N^2 \\
         \end{array}
       \right) \label{eq:totalcij}
\ee
where $N=63$ corresponds to the number of data points of Table \ref{tab:all-data}.
Clearly, this covariance matrix is an oversimplification as it ignores the existing correlations among various data points. Thus, in what follows we consider random variations with reasonable values of nondiagonal elements and identify the effects of these variations on the best fit parameter values and on the tension between these best fit values and \plcdm.
\end{enumerate}
Previous studies have indicated a wide range of tension levels between \plcdm and the growth data depending mainly on the $\fs$ subsample they consider from the data set of Table \ref{tab:all-data}. For example Ref. \cite{Basilakos:2016nyg} finds minimal to no tension with \plcdm while Refs. \cite{Basilakos:2017rgc,Nesseris:2017vor} find about $3\sigma$ tension with \plcdmnospace. Thus a first question we want to address is: ``What is the tension level for the full $\fs$ sample and what are the subsamples that maximize or minimize this tension?''

In Fig. \ref{fig:fullcontour} (left panel) we show the $\Omega_{0m}-\sigma_8$,  likelihood contours obtained from the full data set of Table \ref{tab:all-data} ignoring correlations but including fiducial model corrections. The \plcdm contours are also shown. The tension between the best fit  $\Omega_{0m}-\sigma_8$ and the \plcdm values is at the $5\sigma$ level. The \plcdm parameter values corresponds to higher values of both $\Omega_{0m}$ and $\sigma_8$ indicating stronger clustering than the indication of the actual data. This is also evident in Fig. \ref{fig:fs8z} where the $\fs$ curve corresponding to \plcdm is higher than the majority of the data points. The curve is lower and in better agreement with the full data set for the \wlcdm parameter values that correspond to lower $\Omega_{0m}$ and $\sigma_8$. This weaker clustering, compared with \plcdm preferred by the growth data could be achieved in three ways: by decreasing the value of $\Omega_{0m}$, by decreasing $\sigma_8$ or by decreasing $G_{\rm eff}$ at low redshifts \cite{Nesseris:2017vor}.

\begin{figure*}[ht!]
\centering
\includegraphics[width = 1.\textwidth]{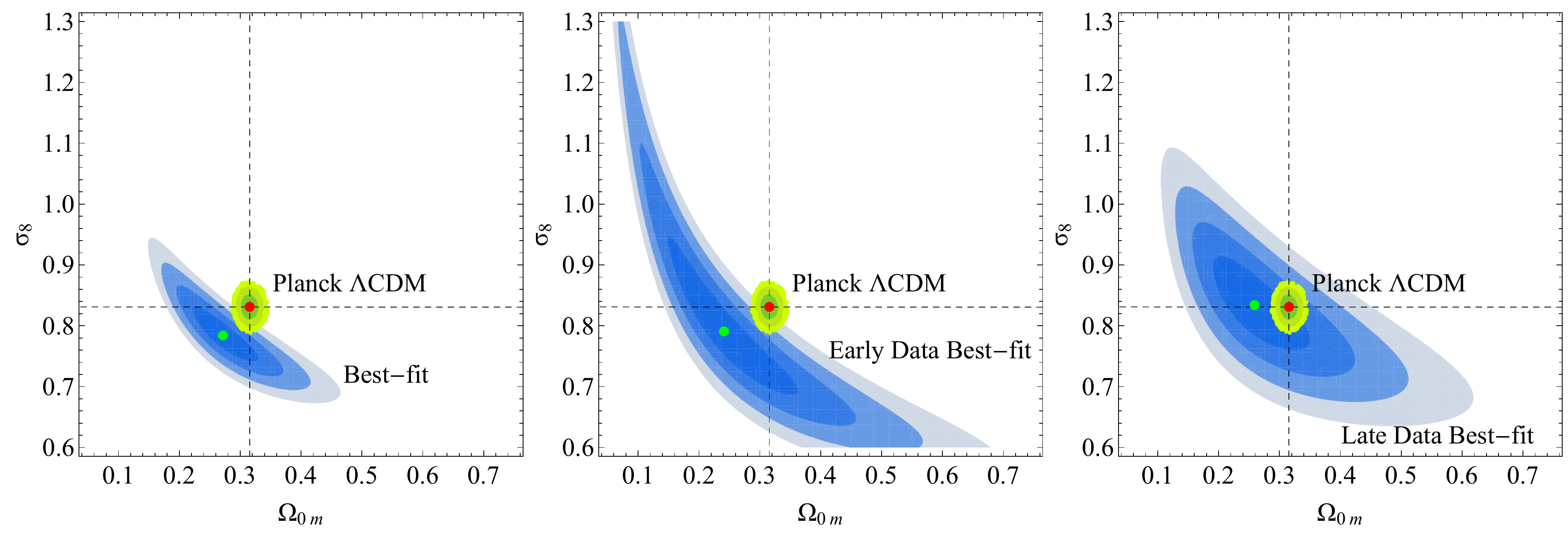}
\caption{Same as Fig. \ref{fig:fullcontour} but with no fiducial cosmology correction. The tension level in all three panels remains approximatelly the same.}
\label{fig:contournofid}
\end{figure*}
The self consistency of the growth data set of Table \ref{tab:all-data} may be tested in several ways. One interesting consistency test is the comparison of the tension level with \plcdm of the early published with the more recently published data. Thus in Figs. \ref{fig:fullcontour} (middle panel) and \ref{fig:fullcontour} (right panel) we show the $\Omega_{0m}-\sigma_8$ likelihood contours obtained using the 20 earliest published data (top 20 points in Table \ref{tab:all-data} where the points are ordered in accordance with time of publication) and the 20 most recently published data (bottom 20 points in Table \ref{tab:all-data}). As shown in Fig. \ref{fig:fullcontour} despite of the increase of the size of the contours due to the smaller number of data, the tension level remains at about $4\sigma$ when the early data are considered. 

In contrast, when the late data are considered (see right panel in Fig. \ref{fig:fullcontour}) the tension level decreases dramatically and the $\sigma$-distance between the best fit $\Omega_{0m}-\sigma_8$ parameters and the corresponding \plcdm parameter values drops below $1\sigma$. This dramatic decrease could be due to following:
\begin{enumerate}[(i)]
\item
The fiducial models considered in early data points that were different from the \plcdm fiducial model assumed in more recent studies. In order to estimate the effects of the assumed fiducial model we reconstruct the contours of Fig. \ref{fig:fullcontour} without implementing the fiducial model correction described by Eq. \eqref{eq:fs8corr}. The new contours are shown in Fig. \ref{fig:contournofid} for the full data set (left panel), for the 20 early data (middle panel) and for the 20 more recent data (right panel). The qualitative feature of the reduced tension for late data remains practically unaffected. Thus, the choice of the fiducial cosmology is not important in identifying the level of tension with \plcdmnospace.

This is also seen by plotting the correction factor $q(z,\Omega_{0m}^{Planck15},\Omega_{0m}')$ as a function of the redshift shown in Fig. \ref{fig:qplot} for various values of $\Omega_{0m}'$. 
\begin{figure}[!h]
\centering
\includegraphics[width = 0.48\textwidth]{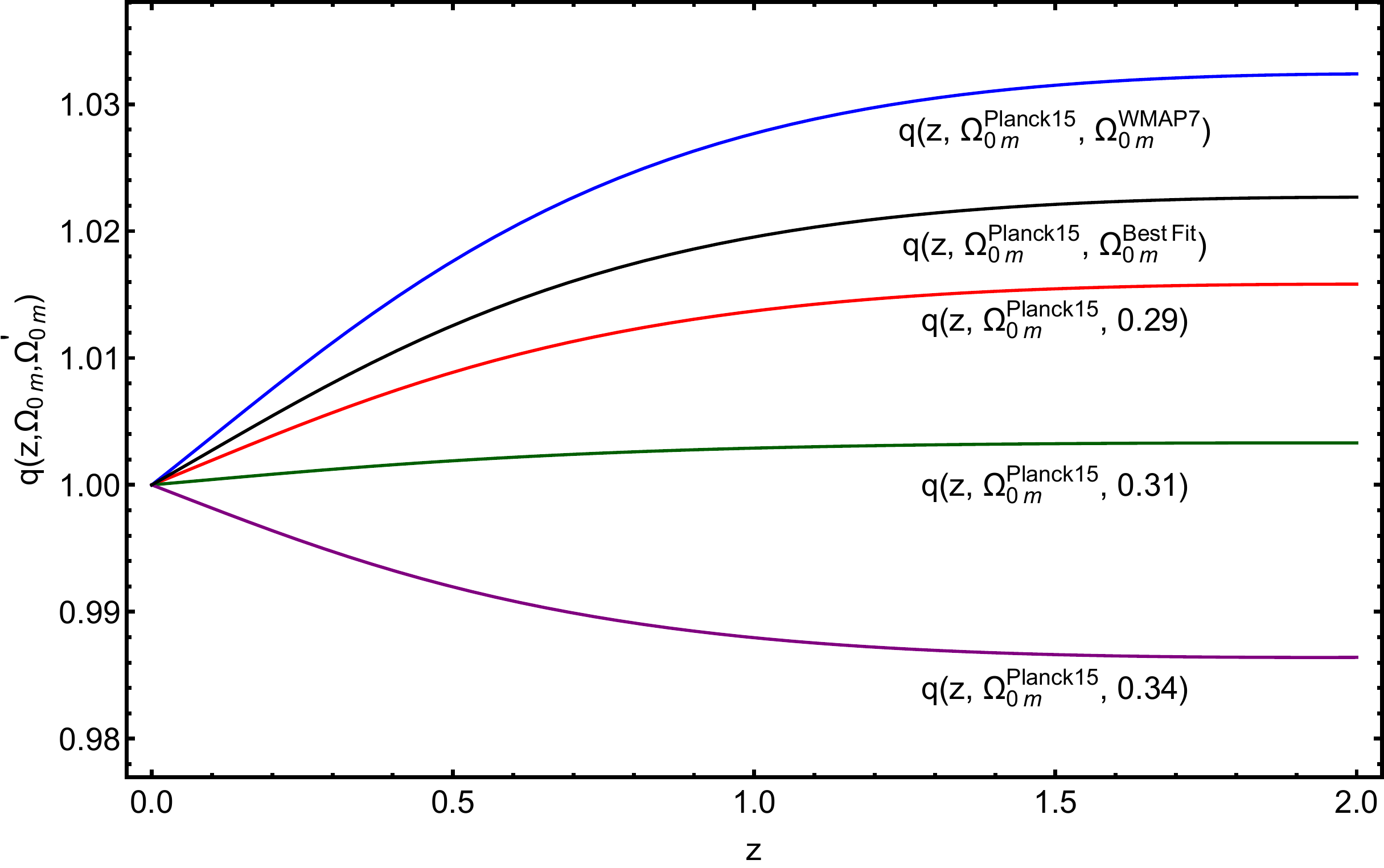}
\caption{The correction factor $q(z,\Omega_{0m}^{Planck15},\Omega_{0m}')$ as a function of the redshift z.}
\label{fig:qplot}
\end{figure}
The cosmological parameters of \wlcdm are chosen as they represent well the fiducial models used for the 20 early $\fs$ data. Clearly, the difference of the correction factor from unity remains less than $3\%$ for redshifts less than 1. This is much less than the  typical level of error bars and explains the reduced role of the fiducial model in determining the tension level of the growth data with the \plcdm parameter values.
\begin{figure*}[ht!]
\centering
\includegraphics[width = 1.\textwidth]{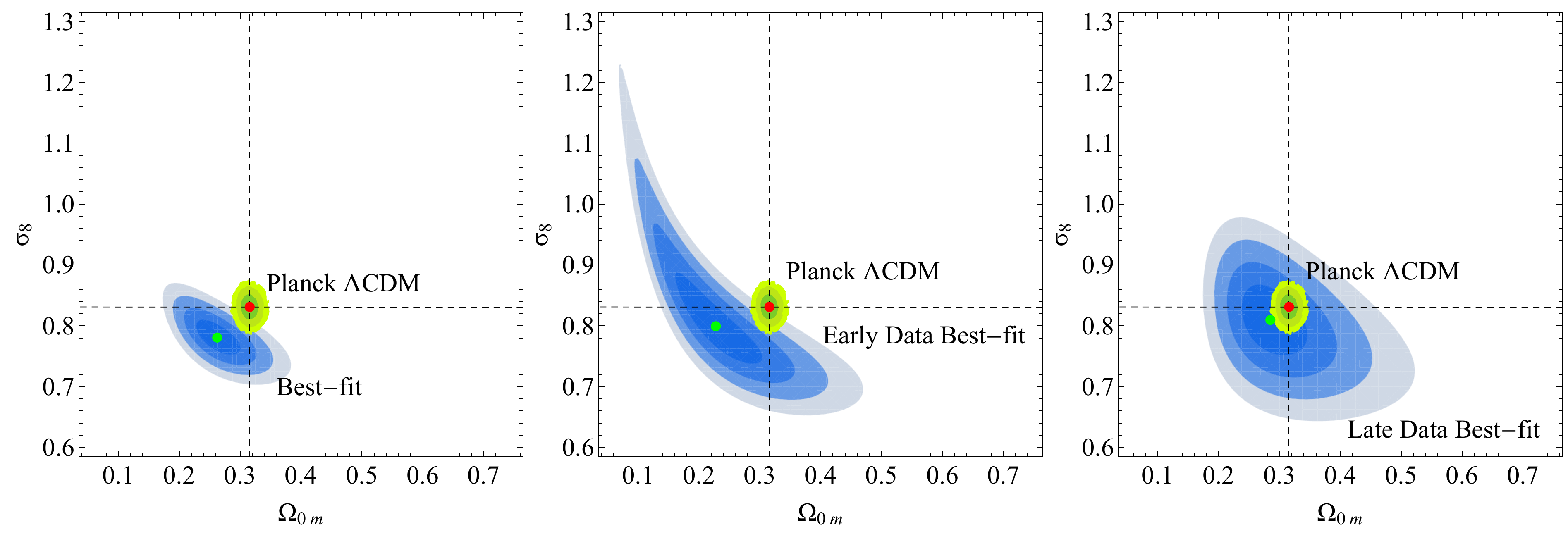}
\caption{Same as Fig. \ref{fig:fullcontour} but with a random covariance among $25\%$ of the data points (assumed to be correlated in pairs). The tension level in all three panels remains approximatelly the same.}
\label{fig:contourcov}
\end{figure*}
\item
The covariance matrix which has been assumed to leave most of the data points uncorrelated. The effects of possible correlations among data points can be estimated by introducing a number of randomly selected nondiagonal elements in the covariance matrix while keeping it symmetric. In this approach we introduce positive correlations in 12 randomly selected pairs of data points (about $20\%$ of the data). The positions of the non-diagonal elements are chosen randomly and the magnitude of the randomly selected covariance matrix element $C_{ij}$ is set to
\be 
C_{ij} =0.5 \sigma_i \sigma_j
\label{cij}
\ee
where $\sigma_i \sigma_j$ are the published $1\sigma$ errors of the data points $i,j$. The coefficient 0.5 is chosen in analogy with the magnitude of the nondiagonal elements of the WiggleZ survey covariance matrix.  The $\Omega_{0m}-\sigma_8$ likelihood contours corresponding to Fig. \ref{fig:fullcontour} with the use of a nontrivial covariance matrix constructed as described above, is shown in Fig. \ref{fig:contourcov}. The qualitative features of Figs. \ref{fig:contourcov} and \ref{fig:fullcontour} remain similar for  the full data set as well as the early data where there is $5\sigma$ tension with the \plcdm parameter values while this tension disappears for the 20 most recently published data points. Thus the introduction of a nontrivial covariance matrix does not change the qualitative conclusions of our analysis which indicate a significant evolution (decrease) of the level of the tension with the time of publication of the $\fs$ data.
\item
Increased redshifts of more recent data points that probe redshift regions where different \lcdm models make similar predictions as shown in Fig. \ref{fig:errorbar_redsh} (bottom panel). This degeneracy is due to matter domination that appears in all viable models at early times. Due to the probe of higher redshifts the more recent data points also have larger error bars a fact that also make them less powerful in distinguishing among different models. The fact of increased redshifts and errorbars for recent data points is demonstrated in what follows. 
\item
Improved methods and reduced systematics may have lead to stronger evidence in favour of the concordance \plcdm cosmological model.
\end{enumerate}
To summarize, the sigma differences for all the cases of contours can be seen in the following Table \ref{tab:sigma}

\begin{table}[h!]
\caption{Sigma differences of the best fit contours from the \plcdm  for Fig. \ref{fig:fullcontour}, Fig. \ref{fig:contournofid} and Fig. \ref{fig:contourcov}.}
\label{tab:sigma}
\begin{centering}
\begin{tabular}{lccc}
 \hhline{====}
  \rule{0pt}{3ex}  
&  Full Dataset & Early Data & Late Data\\
    \hline
Fig. \ref{fig:fullcontour} Contours & $4.97\sigma$ & $3.89\sigma$ & $0.94\sigma$\\
Fig. \ref{fig:contournofid} Contours & $5.44\sigma$ & $4.36\sigma$  & $0.97\sigma$\\
Fig. \ref{fig:contourcov} Contours &$ 4.76\sigma$ & $4.77\sigma$ & $0.37\sigma$\\
\hhline{====}
\end{tabular}
\end{centering}
\end{table}

\newpage
The trend for reduced tension of the growth data with \plcdm may be seen more clearly by plotting the residuals of the data points of Fig. \ref{fig:fs8z} with respect to the \plcdm $\fs$ prediction. These residuals are defined as
\be
\delta f\sigma_8(z_i)\equiv \frac{f\sigma_8(z_i)^{data}-f\sigma_8(z_i)^{Planck15}}{\sigma_i}
\label{eq:resfs8}
\ee
In Fig. \ref{fig:Residual_mov_av} we show these residual data points (with \plcdm fiducial model corrections) ordered with respect to time of publication (top panel) and the corresponding N point moving average (bottom panel) setting $N=20$. The moving average can be defined as
\be
\overline{{f\sigma_8}_j}\equiv \sum_{i=j-N}^j\frac{\delta f\sigma_8(z_i)}{N}
\label{eq:movavdef}
\ee 

\begin{figure}[!h]
\centering
\includegraphics[width = 0.48\textwidth]{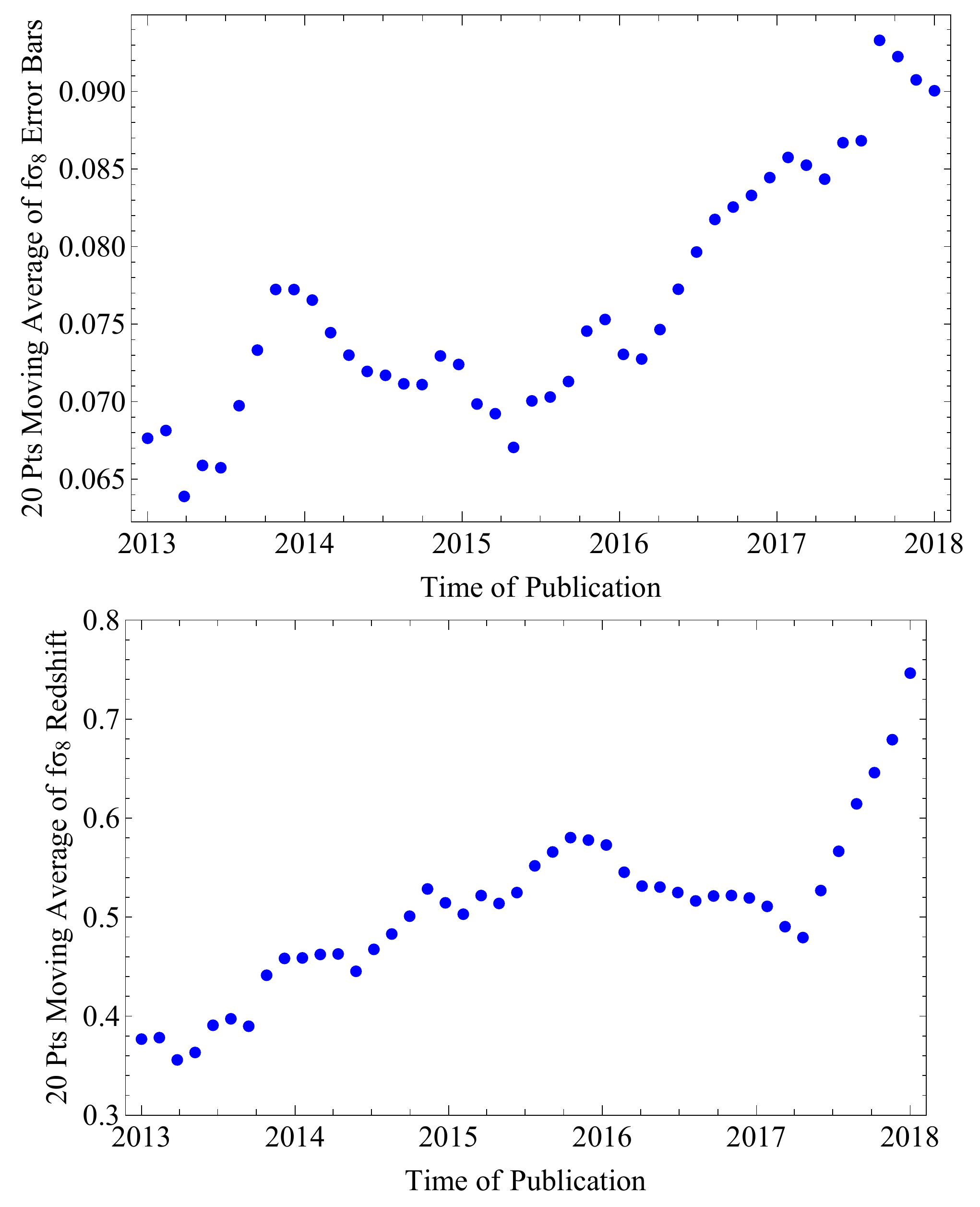}
\caption{\textit{Top panel:} The 20 point moving average of $\fs$ error bars dependence on time of publication. \textit{Bottom panel:}  The 20 point moving average of $\fs$ redshifts dependence on time of publication.}
\label{fig:errorbar_redsh}
\end{figure}

\begin{figure}[!h]
\centering
\includegraphics[width = 0.5\textwidth]{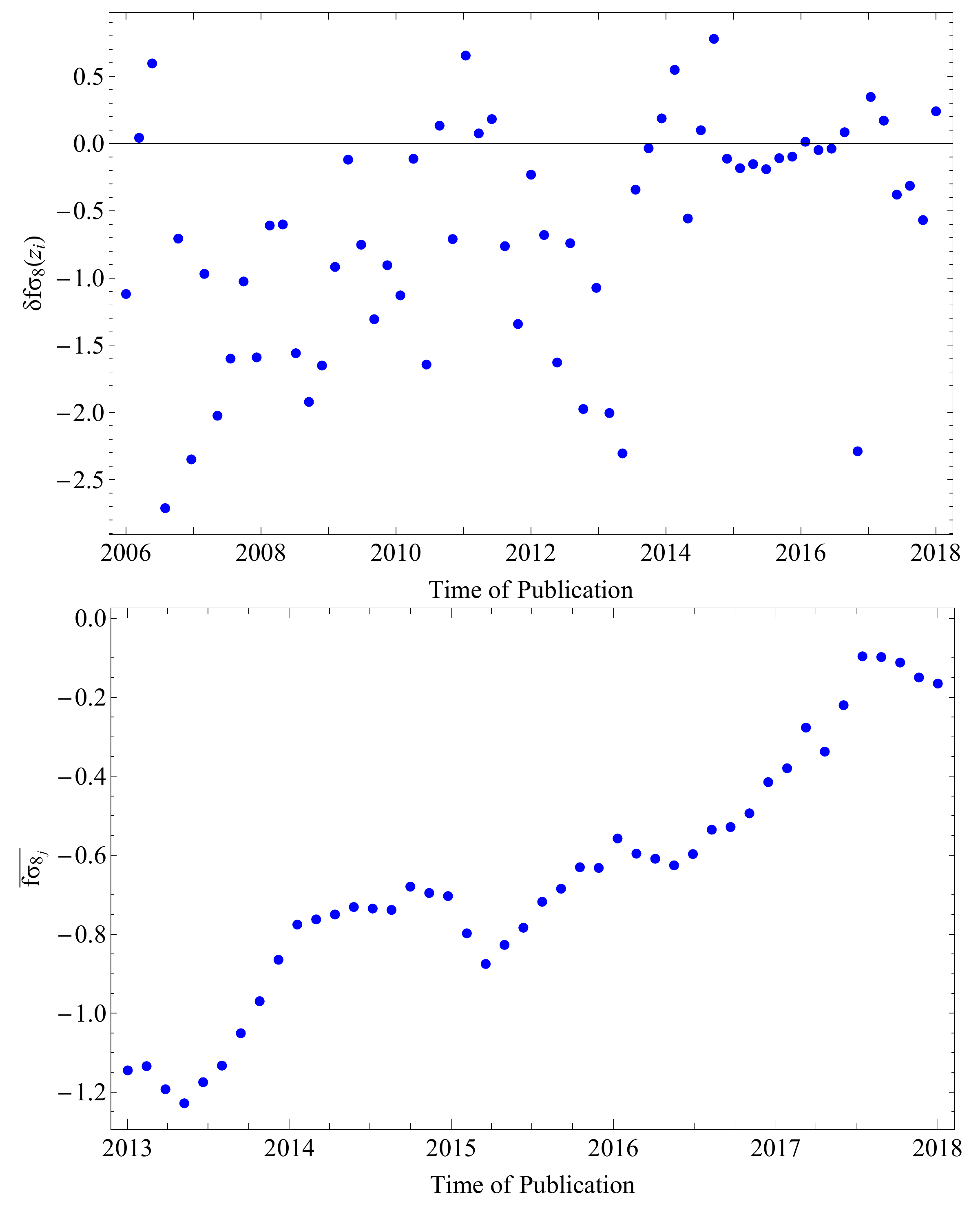}
\caption{\textit{Top panel:} The residual of the data points with \plcdm fiducial model correction based on Eq. \eqref{eq:resfs8}. \textit{Bottom panel:} The 20 points moving average defined by Eq. \eqref{eq:movavdef} with time of publication.}
\label{fig:Residual_mov_av}
\end{figure}

Clearly the consistency of the growth data with \plcdm improves steadily with time of publication. The corresponding moving averages of the error bars and published data redshifts are shown in Fig. \ref{fig:errorbar_redsh} indicating that both the moving average redshift and error bar increase with time of publication (top panel).

The increase of the average data redshift is to be expected due to the improvement of sensitivity of surveys. However, the increased error bars is an unexpected feature and deserves further investigation in view also of the fact that previews studies \cite{Macaulay:2013swa} have indicated that the $\fs$ error bars may be overestimated. 

\begin{figure}[!h]
\centering
\includegraphics[width = 0.48\textwidth]{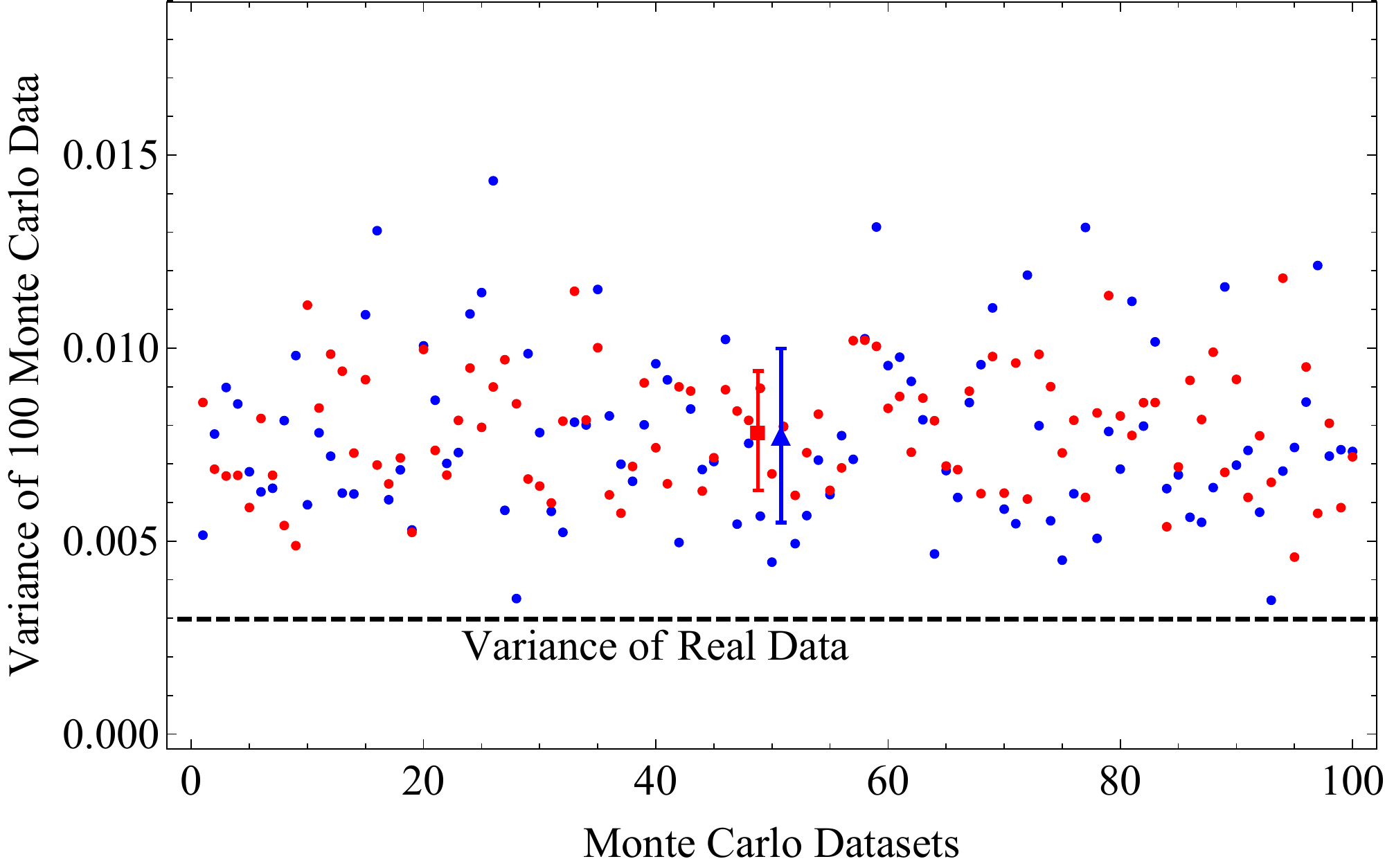}
\caption{The variances of 100 Monte Carlo  data sets. Each red circular point provides the variance of an  Monte Carlo residual data set with uncorrelated data, while each blue point provides the variance of a residual data set with $25\%$ double-counted data points (12 identical pairs of data points). The red square point describes the mean value of the uncorrelated data variances with the standard deviation whereas the blue triangular point is the mean value of the correlated data variances with one standard deviation.}
\label{fig:variance}
\end{figure}

We thus address the following question: ``Are the $\fs$ error bars of Table \ref{tab:all-data} consistent with the spread of the $\fs$ points?'' In order to address this question we compare the variance of the real data $f\sigma_8$ residuals from their best fit \lcdm with the variance of 100 Monte Carlo realizations of the corresponding residual data. In each Monte Carlo realization of the 63 residual data points, each data point is generated randomly from a Gaussian distribution with zero mean and standard deviation equal to the error bar of the real data point. The Monte Carlo variances are shown in Fig. \ref{fig:variance} (100 red dots) along with the variance of the real data residuals (dotted line). The variance of the 100 Monte Carlo residual data sets is $\sigma_{MC}^2=0.0079\pm 0.0015$ while the variance of the real data residuals is $\sigma_{Real Data}^2=0.0030 \pm 0.055$. 

This reduced variance of the real data could be due to either overestimation of the errors of the $\fs$ data of Table \ref{tab:all-data} or due to correlations/double counting in these data. In order to estimate the effects of correlations we introduce artificial double counting in the Monte Carlo data by enforcing $25\%$ of the data points to have an identical corresponding data point is the Monte Carlo $\fs$ data set. The corresponding results after introducing artificial double counting in $25\%$ of the Monte Carlo data can also been seen in Fig. \ref{fig:variance} (blue points). In this case the variance of the Monte Carlo data becomes $\sigma_{MC}^2=0.0077 \pm 0.0023$ which is still significantly larger than the variance of the real data. Thus a moderate level of double-counting is not enough to explain the reduced spread of the real data. This implies that either the error bars of the $\fs$ data are indeed overestimated or that there are systematic effects that prevent the data from having the anticipated from the error bars spread.

\subsection{Implications for modified gravity}

The trend for reduced tension of the growth data with \plcdm with time of publication implies also a trend for reduced indications for evolution of the effective Newton's constant $G_{\rm eff}$. This trend is well parametrized by the parameter $g_a$ of Eq. \eqref{eq:geffansatz}. 

\begin{figure}[!h]
\centering
\includegraphics[width = 0.48\textwidth]{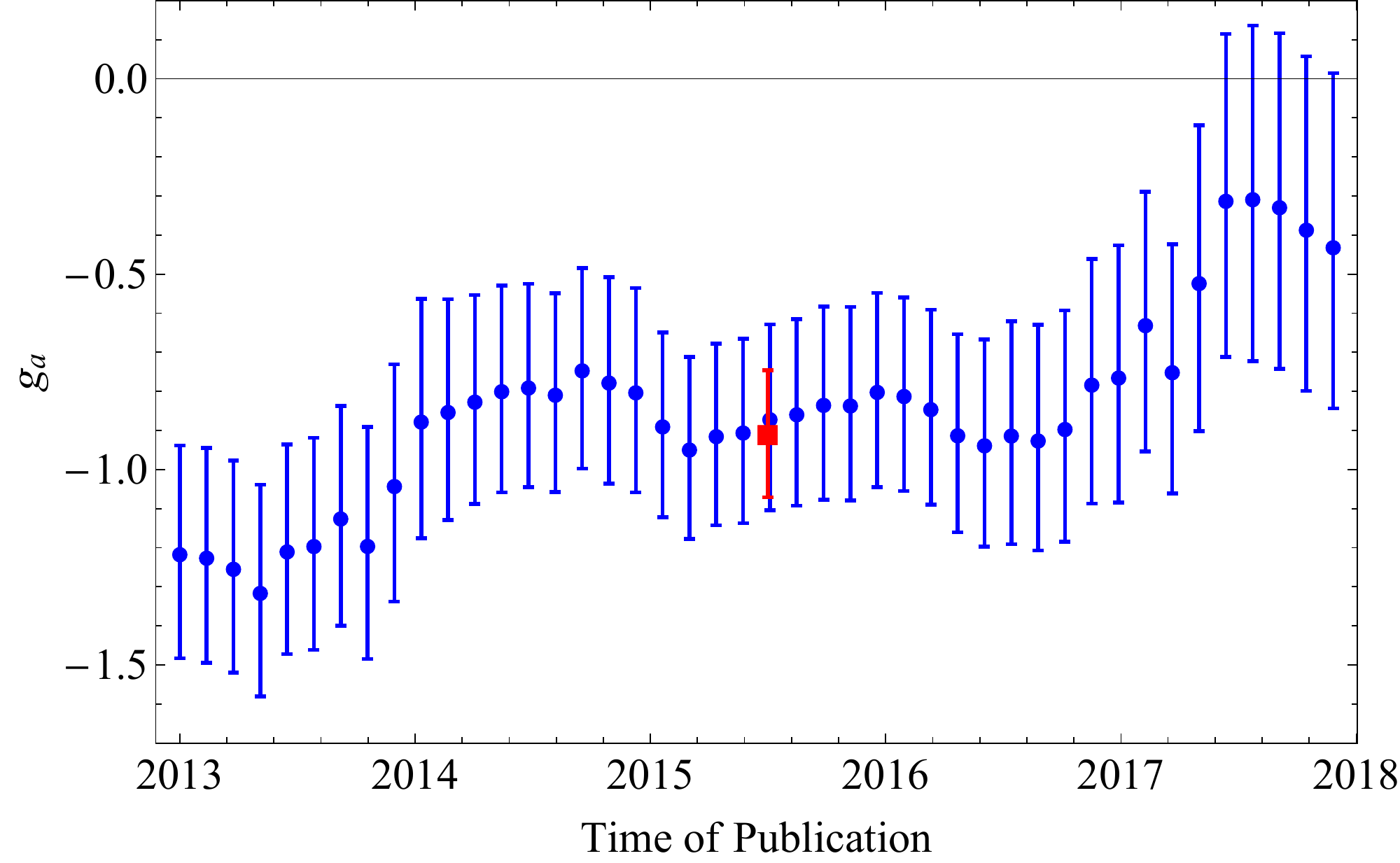}
\caption{The $1\sigma$ range implied for $g_a$ from the full $\fs$ data set. Every blue point corresponds to the best fit $g_a$ obtained from a 20 point $\fs$ subsample starting from the earliest subsample (on the left) to the latest (on the right). The red square point describes the the best fit $g_a$ obtained from the full data set along with the error bar.}
\label{fig:modgrav}
\end{figure}

Assuming a \plcdm background we fit the theoretically predicted $f\sigma_8(z,\Omega_{0m}^{Planck15},\sigma_8^{Planck15},g_a)$ obtained from Eqs. \eqref{eq:odedeltaz} and \eqref{eq:fs8} to the full data set of Table \ref{tab:all-data} as well as to  early and recent subsets in order to identify the evolution of the hints for modified gravity implied by the growth data. In Fig. \ref{fig:modgrav} we show the $1\sigma$ range implied for $g_a$ from the full $\fs$ data set, and for 20 point $\fs$ subsamples starting from the earliest subsample and ending with the most recent subsample.

The $1\sigma$ range for $g_a$ using the full data set of Table \ref{tab:all-data} is $g_a=-0.91\pm 0.17$ (red point). The 20 point subsample best fits  start from $g_a=-1.28^{+0.28}_{-0.26}$ (earliest subsample) which is inconsistent with GR ($g_a=0$) at about $4.5\sigma$ level and ends with the subsample of the 20 most recent data points which imply $g_a=-0.43^{+0.46}_{-0.41}$ which is less than $1\sigma$ away from the GR prediction $g_a=0$.

\section{Conclusion-Discussion}
\label{sec:Section 4}

We have constructed a large $\fs$ growth data set which includes the vast majority (if not all) of the $\fs$ RSD data published to date by several redshift surveys. The data set consists of 63 distinct data points published by different surveys and/or at different times and to our knowledge is the largest $\fs$ compilations that has appeared in the literature so far. Even though this data set is plagued by correlations among data points and possible double counting it is still useful in identifying general trends of the data as well as the sensitivity of the best fit parameters to the fiducial model corrections and to correlations among the data points. Taking various subsamples of the full data set we have demonstrated that the consistency of the published $\fs$ data with \plcdm has improved significantly for the data published during the last 2-3 years. In fact for these data there is currently no tension with the \plcdm in contrast with earlier data published before 2016 which are at about $3-5\sigma$ tension with \plcdmnospace. A partial cause for this reduced tension is the fact that more recent data tend to probe higher redshifts (with higher errorbars) where there is degeneracy among different models due to matter domination. Thus probing redshifts less than one may be a more effective way for distinguishing among different cosmological models.

In addition we have demonstrated that a parametrization of the form of Eq. \eqref{eq:fs8param} provides an excellent fit to the the product $\fs(z)$ obtained from the numerical solution of the Eq. \eqref{eq:odedeltaz} with Eq. \eqref{eq:fs8} in both GR and in modified gravity theories. 

Alternative data sets that directly probe the linear growth rate of density perturbations include weak lensing data (eg. KiDS \cite{Amon:2017lia,Joudaki:2017zdt} or the DES data \cite{Troxel:2017xyo,Baxter:2018kap,Jeffrey:2018cvw,Abbott:2018jhe}) and the Planck Sunyaev-Zeldovich (SZ) cluster counts \cite{Ade:2015fva}. Even though the preferred values of $\sigma_8-\Omega_{0m}$ as obtained from the KiDS weak lensing data and from the Planck SZ cluster counts are in tension with the Planck analysis of primary fluctuations(\plcdmnospace) they are significantly more consistent with the RSD growth data. This fact is demonstrated in Table \ref{tab:sigmaweaklensing} where we show the $\sigma$-distance of the KiDS and Planck cluster $\sigma_8-\Omega_{0m}$ best fits from the RSD data $\sigma_8-\Omega_{0m}$ best fit. 

This Table indicates that the three sets of data that are probing directly the growth rate of cosmological fluctuations (weak lensing, RSD and Planck clusters) are consistent with each other but they are in some tension with the Planck analysis of primary fluctuations which is not as sensitive to the late redshift growth rate of perturbations. This effect could be viewed either as a hint of systematics in the data that probe directly the growth rate of density perturbations or as an early hint of new physics (perhaps of gravitational origin). The detailed investigation of this effect using both early and late weak lensing and cluster number counts data is an interesting extension of this analysis.
\begin{table}[h!]
\caption{Sigma distances of the best fit parameter values of other growth sensitive data sets from the RSD data $\sigma_8-\Omega_{0m}$ best fits.}
\label{tab:sigmaweaklensing}
\begin{centering}
\begin{tabular}{lccc}
  \hhline{====}
  \rule{0pt}{3ex}  
 &  Full RSD  & Early RSD & Late RSD\\
 Dataset &  Data &  Data & Data\\
   \hline
    \rule{0pt}{3ex}  
KiDs Data \cite{Joudaki:2017zdt}  & $1.17\sigma$ & $0.42\sigma$  & $1.50\sigma$ \\
($\Omega_{0m}=0.295^{+0.052}_{-0.087},$  &  &   & \\
$ \sigma_8=0.747^{+0.093}_{-0.125}$)  &  &   &  \\
Planck Clusters Data \cite{Ade:2015fva} &$ 1.21\sigma$ & $1.52\sigma$ & $1.23\sigma$\\
($\Omega_{0m}=0.33 \pm 0.03,$ & & &\\
$ \sigma_8=0.76 \pm 0.03$) & & & \\
 \hhline{====}
\end{tabular}
\end{centering}
\end{table}

Other interesting extensions of the present work include
the search for possible tensions between early and more recently published data in different
data sets including geometric probes (SnIa and BAO)
as well as dynamical growth probes such as weak lensing data. For example as mentioned above, the KiDs
data have indicated significant tension with
\plcdm while this tension is not as strongly supported by other weak lensing data such as the DES data \cite{Troxel:2017xyo,Baxter:2018kap,Jeffrey:2018cvw,Abbott:2018jhe}. 

Finally, our analysis indicates all the $\fs$ subsamples indicate that $G_{eff}$ has higher probability to be decreasing with redshift at low $z$ than to be constant as indicated by GR. Thus it would be interesting to identify those modified gravity models that are consistent with this indication.
\\ \\
\textbf{Numerical Analysis Files}: The numerical files for the reproduction of the figures can be found at \href{http://leandros.physics.uoi.gr/growth-tomography/}{http://leandros.physics.uoi.gr/growth-tomography/}.
\section*{Acknowledgements}
We thank Savvas Nesseris for useful discussions.

\appendix
\section{Fiducial Cosmology Correction}
\label{sec:Appendix_A}

The proper way to homogenize the data set with respect to different fiducial cosmologies would be to recalculate all the $\fs$ data points using the same fiducial cosmology in the construction of the correlation function. This approach is not practical as it would require a recalculation of $\fs(\Omega_{0m},\sigma_8)$ for all parameter values $(\Omega_{0m},\sigma_8)$ for which a value of $\chi^2$ is to be calculated. An alternative approximate approach is the use of correction factors like the one of Eq. \eqref{eq:fs8corr} which are obtained in the context of specific approximations.
Such approaches include the following
\begin{itemize}
\item The fiducial correction in Ref. \cite{Macaulay:2013swa}, used in our analysis through Eq. \eqref{eq:fs8corr}. This correction factor tends to slightly increase the value of the $\fs$ data points when transforming from a \wlcdm fiducial model to a \plcdm model as shown in Fig. 1 of Ref. \cite{Macaulay:2013swa}.
\item The fiducial correction described in Ref. \cite{Alam:2015rsa} where the transformation of $\fs$ from WMAP best fit cosmology \cite{Hinshaw:2012aka} to the Planck best fit cosmology \cite{Ade:2013sjv} is considered. Setting the \wlcdm as the fiducial model and \plcdm as the true cosmology, the relations between the corresponding three dimensional correlation functions taking into account the AP effect is 
\be
\xi_{Planck}(d \ell_\parallel ,d \ell_\perp) =\xi_{fid.}(f_\parallel d \ell_\parallel ,f_\perp d \ell_\perp)\label{eq:corrfunsilv}
\ee
where $f_\parallel=H_{fid.}/H_{planck}$, $f_\perp=D_A^{planck}/D_A^{fid}$.  The corresponding relation between the $\fs$ under specific approximations (e.g. the bias is assumed proportional to $\sigma_8$) may be shown \cite{Alam:2015rsa} to be 
\be
{f\sigma_8}_{Planck} = {f\sigma_8}_{fid.} C \left(\frac{f_\parallel}{f_\perp^2} \right)^{(3/2)} \left(\frac{\sigma_8^{planck}}{\sigma_8^{fid.}} \right)^2\label{eq:growtheqsilv}
\ee
where $C=\int_{k_1}^{k_2} \, dk \sqrt{\frac{P^{fid.}_m}{P^{Planck}_m}}=\int_{k_1}^{k_2} \, dk \sqrt{\frac{P'_m}{P_m}}$. Substituting the definitions of  $f_\parallel$ and $f_\perp$,  Eq. \eqref{eq:fs8corr} takes the following form
\be
q(z,\Omega_{0m},\Omega_{0m}') =C \left(\frac{H'(z) D_A'(z)^2}{H(z) D_A(z)^2} \right)^{3/2} \cdot \left( \frac{\sigma_8}{\sigma_8'} \right)^2\label{eq:qsilvestri}
\ee
Using Eq. \eqref{eq:qsilvestri} for fiducial model correction in our analysis (setting  $C=1$) does not change the trend of reduced tension with \plcdm for the more recent $\fs$ data. However it does reduce significantly the overall tension of \plcdm with the early data.
The new tension levels in the context of the correction factor \eqref{eq:qsilvestri} are shown in Table \ref{tab:sigmasilvestri} 
\begin{table}[h!]
\caption{Sigma differences of the growth data best fit parameter values from the \plcdm under the fiducial correction of Ref. \cite{Alam:2015rsa}.}
\label{tab:sigmasilvestri}
\begin{centering}
\begin{tabular}{l c c c}
 \hhline{====}
  \rule{0pt}{3ex} 
&  Full  & Early & Late \\
& Dataset & Data & Data\\
    \hline
    \rule{0pt}{3ex} 
Correction factor \eqref{eq:qsilvestri} & $2.15\sigma$ & $1.49\sigma$ & $0.86\sigma$\\
No correction & $5.44\sigma$ & $4.36\sigma$  & $0.97\sigma$\\
\eqref{eq:qsilvestri} with random covariance &$ 2.27\sigma$ & $1.15\sigma$ & $0.67\sigma$\\
\hhline{====}
\end{tabular}
\end{centering}
\end{table}
\item 
An alternative fiducial correction factor \cite{Wilson:2016ggz} is written as 
\be
\fs'=\left(\beta+\frac{n}{2} \left(1- \frac{H' \, D_A'}{H \, D_A}\right)\right)b \, \sigma_8\equiv \fs + \fs^{corr} 
\label{eq:wilson}
\ee
where $b$ is the bias, $n$ is the logarithmic derivative of the power spectrum ($n=\frac{d lnP}{dlnk}$).
\end{itemize}
The practical implementation of correction factors \eqref{eq:qsilvestri} and \eqref{eq:wilson} is not as straightforward as the implementation of Eq. \eqref{eq:fs8corr} as the former require information about the power spectrum.

\raggedleft
\bibliography{Bibliography}

\end{document}